\documentclass[a4paper,11pt]{article}
\pdfoutput=1
\usepackage{jcappub}
\usepackage{afterpage,booktabs}
\usepackage{graphicx,subfig}
\usepackage{amssymb,amsmath,bm}
\usepackage{mathtools}
\usepackage[usenames,dvipsnames]{xcolor}
\usepackage[capitalise]{cleveref}
\usepackage{expl3,xstring}
\usepackage{dcolumn}
\usepackage{bm}
\usepackage{color}
\usepackage{breqn}
\usepackage{multirow}
\usepackage[normalem]{ulem}
\usepackage{tabu}
\usepackage{xcolor}
\usepackage{fontawesome}

\definecolor{earlyde}{RGB}{255, 192, 203}
\definecolor{unphysical}{RGB}{106,90,205}
\definecolor{nomatterdom}{RGB}{218,112,214}
\definecolor{viable}{RGB}{0,0,128}
\definecolor{star}{RGB}{255,165,0}
\definecolor{royalgreen}{RGB}{19,120,7}

\newcommand{\HiCOLA}{\texttt{Hi-COLA\;}}

\newcommand{\lcdm}{$\Lambda$CDM}
\newcommand{\mplanck}{M_{\rm {Pl}}}

\newcommand{\ttau}{\tilde{\tau}}

\newcommand{\NP}[0]{N_{\rm part}^{1/3}}
\newcommand{\MP}[0]{M_{\rm part}}
\newcommand{\NM}[0]{N_{\rm mesh}^{1/3}}
\newcommand{\NS}[0]{N_{\rm step}}
\newcommand{\LF}[0]{\ell_{\rm Force}}
\newcommand{\Msun}{\, h^{-1} \,  M_{\odot}}
\newcommand{\hompc}{\,h\,{\rm Mpc}^{-1}}
\newcommand{\mpcoh}{\,h^{-1}\,{\rm Mpc}}

\title{\boldmath K-mouflage at high $k$: extending the reach of \HiCOLA}

\author{Ashim Sen Gupta$^1$,}
\author{Bartolomeo Fiorini$^2$,}
\author{Tessa Baker$^2$}

\affiliation{$^1$Astronomy Unit, Queen Mary University of London, Mile End Road, London E1 4NS, UK}
\affiliation{$^2$Institute of Cosmology and Gravitation, University of Portsmouth, \\
Burnaby Road, Portsmouth PO1 3FX, UK}

\emailAdd{a.sengupta@qmul.ac.uk}
\emailAdd{bartolomeo.fiorini@port.ac.uk}
\emailAdd{tessa.baker@port.ac.uk}

\date{\today}

\abstract{The \href{https://github.com/Hi-COLACode/Hi-COLA}{\faicon{github} \HiCOLA{}} code is an efficient dark matter simulation suite that flexibly handles the Horndeski family of modified gravity models. In this work we extend the scope of \HiCOLA \ to accommodate Horndeski theories with K-mouflage screening, allowing for the computation of matter power spectra in the non-linear regime in these models. We explore the boost of the dark matter power spectrum relative to GR-$\Lambda$CDM in K-mouflage gravity, and also discuss how large-scale structure computations change between the Einstein and Jordan frames. A dissection of the relative contributions of the modified background, linear growth, fifth force, and the conformal factor (a new inclusion to \HiCOLA) to the boost factor is presented. The ability of \HiCOLA \ to run with general Horndeski models and multiple screening mechanisms makes it an ideal tool for testing gravity with upcoming galaxy survey data.
}

\begin{document}
\maketitle


\section{Introduction}

\subsection{Motivation}

Upcoming surveys such as LSST~\cite{LSST}, Euclid~\cite{Euclid} and DESI~\cite{DESI} will provide key data to test gravity by observing large-scale structure (LSS) on non-linear scales. This will prove crucial in the efforts to constrain the space of modified gravity theories, particularly those with screening mechanisms. These mechanisms, which manifest on scales beyond the reaches of traditional perturbative analysis, enable modified gravity theories to behave like General Relativity (GR) on Solar System scales. It is for this reason that the non-linear scales of LSS formation are promising, as one can hope to probe the transition from unscreened to screened regimes\footnote{It should be noted that accurately modelling structure formation on small scales will ultimately necessitate detailed bias models to connect the matter density field to the observed galaxy field, and the inclusion of baryonic effects if one seeks to probe beyond $\mathcal{O}(1)$ Mpc/h.}. Smaller-scale results are also less affected by cosmic variance, owing to the higher realisations of modes. In this way, non-linear LSS formation can serve as a complement to large-scale studies~\cite{Zuntz:2011aq,Baker:2012zs,Simpson:2012ra,Leonard:2015hha} and push the precision of constraints.

To take advantage of this forthcoming data requires the accurate modelling of non-linear (modified gravity) structure formation. The traditional approach is to employ $N$-body codes, e.g.~\cite{Schmidt:2009sg,Schmidt:2009sv,Li2011coupledscalar,Zhao2011fRPM,Llinares:2013jza,Puchwein:2013lza}, for this task, but these are computationally and temporally expensive to run. Thus, there is the added requirement for \textit{rapid} modelling of LSS formation beyond GR-$\Lambda$CDM. The Horndeski-in-COLA  (\href{https://github.com/Hi-COLACode/Hi-COLA}{\faicon{github} \HiCOLA{}}) code was developed to address these needs. Owing to the vastness of the modified gravity theory space, the code was designed to tackle as much of the canonical scalar-tensor space of modified gravity theories as possible, embodied in the Horndeski family. \HiCOLA was first introduced in \cite{Wright:2022krq}, and an overview of the code is provided in Section~\ref{sec:HiCOLA}.
Contemporaries of \HiCOLA{} include codes such as \texttt{ReACT}~\cite{Atayde:2024tnr,Bose:2020wch}, \texttt{MG-GLAM}~\cite{hernandez-aguayo_fast_2022}, \texttt{MG-evolution}~\cite{Hassani:2020rxd} and, in particular, \texttt{COLA-FML}~\cite{Brando:2023fzu} and \texttt{FML}\footnote{As noted in the documentation for \texttt{MG-PICOLA}, \href{https://github.com/HAWinther/FML}{\texttt{FML}} is its actively-developed successor written in \texttt{C++}.}~\cite{Winther:2017jof}, which also adopt the COLA simulation approach. With such a variety of rapid simulation tools, this marks an exciting time for constraining modified gravity theories using LSS observations.

In \cite{Wright:2022krq} we focused on Horndeski theories with Vainshtein screening, which is one the screening mechanisms present in the Horndeski class, alongside, for example, Chameleon and K-mouflage screening~\cite{Khoury:2003aq, Brax:2015pka,kobayashi_horndeski_2019}. As the ultimate aim for \HiCOLA is to address as much of the Horndeski class as possible, in this paper we now extend the scope of \HiCOLA to models with K-mouflage screening. In so doing, we establish the limitations of \HiCOLA's initial release, as the next-to-leading order perturbative calculations used to derive the fifth force in \cite{Wright:2022krq} do not apply to K-mouflage. In Section~\ref{sec:Kmou_vs_Vainshtein}, we will see that K-mouflage theories must be handled on a case-by-case basis depending on the form of the kinetic Horndeski function [$K$ in eq.~\eqref{eq:rH_action}]. This is unlike the Vainshtein case, where a single expression for the fifth force applies to all members of the subclass of Horndeski theories with Vainshtein screening. 

K-mouflage has often been studied theoretically in the Einstein frame, but \HiCOLA operates naturally in the Jordan frame, where matter is minimally coupled to the metric. For this reason we review the differences between the Einstein and Jordan frames on background quantities and the power spectrum in Section~\ref{sec:kmou_Jordan}, and show corresponding results in Section~\ref{sec:power_spectra}.  This is important as some regularly-used quantities of interest are \textit{not} frame-independent observables~\cite{francfort_cosmological_2019}, and this knowledge is crucial in correctly comparing \HiCOLA's Jordan-frame results with the Einstein-frame results for K-mouflage in the literature~\cite{brax_K-mouflage_2014_LSS,hernandez-aguayo_fast_2022}.

The inclusion of K-mouflage theories in \HiCOLA marks a key step in the prospects of testing K-mouflage gravity with data. Alongside the recent implementations of K-mouflage in \texttt{MG-GLAM}~\cite{hernandez-aguayo_fast_2022} and \texttt{ReACT}~\cite{Bose:2020wch,Atayde:2024tnr}, this wealth of tools can be used to generate sufficient K-mouflage clustering data for the training of emulators. These emulators will ultimately enable constraints on the K-mouflage theory space using Stage IV galaxy survey data. We note that, to our knowledge, \HiCOLA presents the first simulation code for K-mouflage in the Jordan frame. As K-mouflage is one of the simplest models beyond the Vainshtein subclass of theories, this is a natural next step in narrowing down the viable set of theories in the broader Horndeski class.

It should be noted that validation of results will not be a focus of this paper, and will instead be covered in~\cite{Bose:2024qbw}.

\subsection{Notation}

\begin{itemize}
    \item Unless specified otherwise, tildes ($\sim$) or an E in the subscript denote quantities in the Einstein frame. The lack of a tilde, or a J in the subscript implies a Jordan-frame quantity.
    \item The background value of the scalar field is explicitly referred to by $\bar{\phi}$. For simplicity, as most sections of this paper refer to the background value, $\phi$ will be used in place of $\bar{\phi}$. Exceptions to this notation, as in Section~\ref{sec:Kmou_vs_Vainshtein}, will be indicated where relevant.
    \item The massless scalar field $\hat{\phi}$ is defined by $\phi = M_{\phi} \hat{\phi}$ (where $M_{\phi}$ is the mass scale of the scalar field). As most sections use the massless scalar field, once again for simplicity $\phi$ will be used to refer to it, rather than $\hat{\phi}$.
    \item Over-dots denote derivatives with respect to coordinate time, i.e. $\dot{f} \equiv \frac{df}{dt}$.
    \item Primes ($'$) denote derivatives with respect to the logarithm of the scale factor, i.e. $f' \equiv \frac{df}{d(\ln a)}$.
    \item Subscripts with (a comma followed by) $\tau$ denote derivatives with respect to conformal time, i.e. $f_{1,\tau} = \frac{df_1}{d\tau}$.
    \item Quantities that depend on time, with a 0 in the subscript are evaluated today, i.e. $f_0 \equiv f(a=1)$.
    \item $X,\phi$ in subscripts imply derivatives with respect to them, i.e. $f_X \equiv \frac{df}{dX}$.
    \item We will assume a system of units where $c=1$.
\end{itemize}

\subsection{Structure of this paper}

This paper begins with a summary of the \HiCOLA{} code in Section~\ref{sec:HiCOLA}. Here we describe the overall layout of the code, and the primary quantities it computes that enable the incorporation of fifth force effects in the simulated clustering. 

In Section~\ref{sec:kmou_einstein} we review the definition of K-mouflage in the Einstein frame, as is typical in much of the literature, and present the background equations and linear growth equation. 

In Section~\ref{sec:kmou_Jordan}, we examine K-mouflage from the Jordan frame, establishing the transformation expressions between the frames and enabling the inclusion of K-mouflage in \HiCOLA, as it is a Jordan-frame code. We show the results for expansion history and linear growth generated using \HiCOLA, comparing Einstein and Jordan frame results. 

In Section~\ref{sec:Kmou_vs_Vainshtein}, we explain why it is not possible to obtain a fully general fifth force expression that applies to all K-mouflage theories. We justify the expressions we will use to implement the particular K-mouflage model of Section~\ref{sec:kmou_einstein} in \HiCOLA{}. A reader primarily interested in simulation results (and less so the theoretical underpinnings of the K-mouflage force) may wish to skip this section on a first reading.

In Section~\ref{sec:power_spectra} we present the non-linear matter power spectra for K-mouflage gravity produced by \HiCOLA. We discuss the aspects of K-mouflage that influence the shape of the boost (the ratio with respect to GR-\lcdm{}), and also discuss the effect of the K-mouflage model.

In Section~\ref{sec:convergence} we perform convergence tests for the \HiCOLA{} K-mouflage gravity results. 

We conclude in Section~\ref{sec:conclusions}.

\section{The {\HiCOLA} code}
\label{sec:HiCOLA}

The \HiCOLA code \cite{Wright:2022krq} was developed with the goal of diversifying the range of modified gravity models that can be studied on non-linear scales. \HiCOLA is an extension of the COLA solver included in the \texttt{FML}\footnote{\url{https://fml.wintherscoming.no}} library by Hans Winther. By their nature, COLA solvers -- those implementing the COmoving Lagrangian Acceleration method \cite{OriginalCOLA,2015arXiv150207751T} -- are efficient and lightweight simulations. They hybridise Lagrangian Perturbation Theory (LPT) calculations for large cosmological scales with a particle-mesh {\it N}-body solver on small scales. Effectively the LPT gives the first- or second-order displacement of the particles, and the {\it N}-body solver computes adjustments to particle positions (and velocities) that are beyond a tractable perturbative regime.

\HiCOLA inherits the speed of COLA solvers, but adds a set of modifications adapted to encompass a broad range of gravity and dark energy models. Specifically, \HiCOLA has been designed to simulate the \textit{reduced Horndeski} family of models. Horndeski gravity \cite{Horndeski1, Horndeski2, Horndeski3, kobayashi_horndeski_2019} is the most general description of gravity theories with one new scalar degree of freedom (in addition to the usual metric of GR) and second-order equations of motion, thereby encompassing much of the well-studied theory space such as quintessence, Jordan Brans-Dicke, f(R) gravity, DGP gravity and many more theories. The additional word `reduced' here implies that we focus on the subset of Horndeski gravity in which gravitational waves travel luminally, consistent with observations of multimessenger event GW170817 and its electromagnetic counterpart GRB170817a~\cite{LIGOScientific:2017ync, LIGOScientific:2017vwq}. The action for reduced Horndeski gravity is given by \cite{Lombriser:2016zfz, Baker:2017hug, Ezquiaga:2017ekz,Creminelli:2017sry}:
\begin{equation}
\label{eq:rH_action}
S = \int d^4 x \sqrt{-g}\left( G_4(\phi) R + K_J(\phi, X) - G_3(\phi, X) \Box \phi  - \mplanck^2 \Lambda + \mathcal{L}_m\left( \gamma_{\{m\}},g_{\mu \nu} \right) \right)\,.
\end{equation}
Here $G_3$, $G_4$ and $K_J$ are free functions of a scalar field $\phi$ and its kinetic term $X=-\partial_\mu\phi\partial^\mu\phi/2$, which in principle can take any functional form (though not all choices will lead to viable cosmologies).

\HiCOLA consists of two main components: 
\begin{itemize}
    \item A front-end theory module. This receives an input gravity model from the user as symbolic expressions for $G_3$, $G_4$ and $K_J$. The relevant Friedmann equations are formed symbolically and solved in a few seconds, along with a few other relevant quantities that can be pre-computed outside a COLA simulation.
    \item The back-end COLA solver. The expansion history (and other quantities) pre-computed by the front-end are fed into the full COLA simulation, from which snapshots and dark matter power spectra can be extracted in the usual manner. Simulations with typical settings \cite{izard_ice-cola_2016} require as little as $\sim 300$ CPUh per billion particles to run.
\end{itemize}
It is useful to break down explicitly the ways in which \HiCOLA accounts for the main effects of Horndeski gravity on non-linear structure:
\begin{enumerate}
    \item The expansion history computed by the front-end may differ from \lcdm \ or a $w_0-w_a$ parameterisation;
    \item The linear growth of structure may differ from \lcdm\footnote{COLA solvers additionally make use of second-order modified growth factors in their LPT calculations. However, in previous studies \cite{Valogiannis:2016ane, Wright:2022krq} these have been found to have negligible effects, so they are currently computed with \lcdm{} solutions in \HiCOLA.};
    \item The effective gravitational constant may differ from Newton's constant in a redshift-dependent manner: $G_N\rightarrow G_{\rm eff}(z)$;
    \item The forces between particles in the COLA simulation differ from Newtonian gravity. These modified forces can be schematically represented as:
    \begin{equation}
\label{eq:fifth_force}
    F_{\textrm{tot}} = F_N\, \frac{G_{\rm eff}(z)}{G_N}\left[1+\beta(z) S(z,\delta_M)\right].
\end{equation}
Here $F_N$ is the standard Newtonian force; we see the effective gravitational constant acting here is modified to be $G_{\rm eff}$, as described above. Inside the square brackets is the additional `fifth force' arising from the scalar field, described by two factors. $\beta(z)$, which we call the \textbf{coupling}, is an enhancement dependent only on redshift. On linear scales (given by $S\rightarrow 1$), this gives the linear force enhancement affecting linear growth rates etc. The expression of $\beta$ in terms of the Horndeski functions can be found in section 3 of \cite{Wright:2022krq}.

The second component $S(z, \delta_M)$, the \textbf{screening factor}, is responsible for the suppression of modified forces in certain environments within the simulations, i.e. screening effects. We implement a spherical screening factor approximation as put forwards in \cite{Winther:2015wla}, which avoids the need to solve the full scalar field equation of motion within the simulation (which would be computationally expensive). We use the argument $\delta_M$ here to indicate that $S$ is a function of the density field in the simulation in some form; we flag that for different screening mechanisms the gravitational potential or its derivatives may be the quantity appearing explicitly. The derivation of $S(z, \delta_M)$ for Vainshtein screening in reduced Horndeski gravity can also be found in section 3 of \cite{Wright:2022krq}.
\end{enumerate}
In this work, we will see all four of these effects come into play in the context of K-mouflage gravity. 

We highlight the unique aspect of \HiCOLA is that it consistently implements all of the modifications 1)-4) originating from a single Lagrangian. In particular, the modified expansion history produces significant effects on both linear and non-linear scales; this is not captured by codes which implement modified forces on top of a \lcdm{}-like expansion history.

\section{K-mouflage Model}\label{sec:kmou_einstein}

\subsection{Action}
K-mouflage is typically defined in the Einstein frame \cite{brax_K-mouflage_2014_LSS}, where the action takes the form 

\begin{equation}
\label{eq:kmou_action}
    S = \int d^4x \sqrt{-\tilde{g}} \left( \frac{M_{\textrm{pl}}}{2} \tilde{R} + \mathcal{L}_{\phi}(\phi) + \mathcal{L}_m( \gamma{\{m\}}, A^2(\phi)\tilde{g}_{\mu \nu}) \right) . 
\end{equation}
$\tilde{R}$ is the standard Einstein-Hilbert term, giving rise to the Einstein field equations. $\phi$ is the dimensionless scalar field, with its contributions contained in $\mathcal{L}_{\phi}$. $\mathcal{L}_m$ is the matter Lagrangian, with matter fields represented collectively by $\gamma{\{m\}}$, and $A(\phi)$ introduces a non-minimal coupling between the scalar field and matter. In the Einstein frame, matter is coupled to $A^2(\phi) \tilde{g}_{\mu\nu} \equiv g_{\mu \nu}$, which is the Jordan frame metric. One can apply a conformal transformation $\tilde{g}_{\mu \nu} \rightarrow A^2 g_{\mu \nu}$ to obtain eq.~\eqref{eq:kmou_action} in the Jordan frame, where matter is minimally coupled to the Jordan-frame metric $g_{\mu \nu}$, but instead there is a non-minimal coupling between the metric and the scalar field. Differences between the Jordan and Einstein frames are discussed further in Section~\ref{sec:kmou_Jordan}.

Following the conventions of the \HiCOLA{} code \cite{Wright:2022krq}, we will pull out explicit mass scales wherever possible, such that general functions appearing in the Lagrangian are always dimensionless. We will also assume that the mass scale of the scalar field, $M_{\phi}$, is of Planckian scale, i.e. $M_{\phi} =\mplanck$, so that the scalar field terms are of roughly the same order significance as the Einstein-Hilbert term. We define the massless scalar field by $\phi = M_{\phi} \hat{\phi}$. However, for simplicity, we will not use the new $\hat{\phi}$ notation for the massless scalar field and instead use $\phi$ directly.

For K-mouflage we have that
\begin{equation}
\label{eq:kmou_lphi}
    \mathcal{L}_{\phi} = M_K^2 H_0^2 K(X),
\end{equation}
where $K$ is a non-standard kinetic term for the scalar field, which is a function of the dimensionless canonical kinetic term $X$ defined below eq.~\eqref{eq:rH_action}. Using $E \coloneqq H/H_0$ to denote the dimensionless Hubble rate, on a cosmological background, $X= E^2 \phi^{\prime 2}/(2\lambda^2)$. We arrive at this \textit{background-level} expression for $X$ by restricting ourselves to homogeneous and isotropic spacetimes, and by changing variables from coordinate time $t$ to $\ln a$. We follow \cite{hernandez-aguayo_fast_2022}, adopting the convention that $\lambda$ is the ratio of the mass scale of the kinetic term $K$, i.e. $M_K$, to the Planck mass. That is, $\lambda = M_K / M_{\textrm{Pl}}$.

The mass scale of $K$ is typically set to values such that K-mouflage contributions to the Friedmann equation are significant relative to matter energy densities.
This ensures the theory is cosmologically relevant~\cite{brax_K-mouflage_2014_LSS}. The parameter $\lambda$ can be adjusted to tune the value of $H_0$ obtained when solving the cosmological expansion history.

Following the choice made in \cite{brax_K-mouflage_2014_background} for the functional form of $K$, and the further attention this model has received \cite{hernandez-aguayo_fast_2022}, we choose the same form. This is shown below in eq.~\eqref{eq:kmou_K}: 
\begin{align}
\label{eq:kmou_K}
    K = -1 + X + K_0 X^n.
\end{align}

The $-1$ term ensures that the vacuum energy in the low excitation limit of $\phi$ is positive, corresponding to an effective cosmological constant. Additionally, we also choose to to follow \cite{brax_K-mouflage_2014_background} in defining an exponential conformal factor: 
\begin{equation}\label{eq:conformal_factor}
    A = \exp\left(\beta_K \phi \right),
\end{equation}
where $\phi$ is the \textit{massless} scalar field. As explained in \cite{Barreira:2015_kmou_solar}, $\beta_K^{\prime}$ is weakly constrained by existing observations. Therefore, eq.~\eqref{eq:conformal_factor} is tantamount to making the simplest choice for the parameter $\beta_K$, featuring as a constant in the linear fifth force modification to the gravitational force (see Section~\ref{sec:fifth_force_einstein}).

The set of parameter constants for the K-mouflage model studied here are $\{n, K_0, \beta_K,\lambda\}$, where we remind the reader that the meaning of $\beta_K$ is made clear in eq.~\eqref{eq:kmou_force} and that $\lambda$ is defined below eq.~\eqref{eq:kmou_lphi}. $\lambda$ will be adjusted to ensure $H_0$ in this model matches values consistent with the Planck 2015 cosmological release~\cite{ade_planck_2016}, matching the approach of~\cite{hernandez-aguayo_fast_2022}.

$n, K_0, \beta_K$ will be varied, and their effects  on matter clustering discussed in Sections~\ref{sec:evolving_k_b} and~\ref{sec:evolving_n}.

\subsection{Background equations}
\label{sec:kmou_bg_einstein}

We first remind readers that we have simplified notation by denoting the background value of the scalar field by $\phi$. The homogeneous, isotropic background equations (assuming dust-matter) for K-mouflage theories  can be depicted in Friedmann-like form as \cite{brax_K-mouflage_2014_background,hernandez-aguayo_fast_2022}:

\begin{align}
    3\mplanck^2 H^2 &= \rho_E + \rho_{\phi}\label{eq:kmou_friedmann},\\
    -2\mplanck^2 \dot{H} &= \rho_E + \rho_{\phi} + P_{\phi} + P_r\label{eq:kmou_raychaudhuri}.
\end{align}
Here $\rho_E$ is the total energy density of cold dark matter, baryons and radiation, and $P_r$ the pressure due to radiation, in the Einstein frame. The effective energy density and pressure of the scalar field are given by (where recall $K_X = \frac{\partial K}{\partial X}$):

\begin{align}
    \rho_{\phi} &= -M_K^4 K + M_{\textrm{Pl}}^2 \dot{\phi}^2 K_X \label{eq:kmou_rhophi},\\
    P_{\phi} &= M_K^4 K.
\end{align}
 Meanwhile the Klein-Gordon equation for the background component of the scalar field is:
\begin{align}
\label{eq:kmou_kg}
    \mplanck \ddot{\phi} \left( K_X + \frac{\mplanck^2 \dot{\phi}^2}{M_K^4} K_{XX}\right) + 3\mplanck H \dot{\phi} K_X = - \frac{\rho_E}{A}\frac{dA}{d\phi}.
\end{align}

The background solutions in this work were obtained by integrating eq.~\eqref{eq:kmou_kg} for a given set of initial conditions for $\{\phi, \dot{\phi}\}$ to find values for $\dot{\phi}$ and  $\phi$ at the next timestep, and the use of the Friedmann-like equation [eq.~\eqref{eq:kmou_friedmann}] to find $H$. The resultant quantities were then carried over into the subsequent timestep and the process repeated. We found that algebraically solving eq.~\eqref{eq:kmou_friedmann} to find $H$, as opposed to integrating eq.~\eqref{eq:kmou_raychaudhuri} as done for the Cubic Galileon in \cite{Wright:2022krq}, had greater stability for K-mouflage background solutions.

A representative example of the resulting expansion history one finds for K-mouflage, relative to \lcdm, is shown by the orange curve in the left panel of Fig.~\ref{fig:kmou_hubble_omega}.

\subsection{Linear Theory}
\label{sec:linear_theory_einstein}
In the Einstein frame, the growth equation for the K-mouflage model reads:
\begin{align}
&\tilde{D}_{1,\ttau \ttau}+\left[\mathcal{H}_E+\frac{\mathrm{d} \ln A(\varphi)}{\mathrm{d} \phi} \varphi_{\ttau}\right] \tilde{D}_{1,\ttau}-\frac{3}{2} \Omega_{\rm m}(a_E) {H}_0^2 a_E^2 A(\bar{\varphi})\mu \tilde{D}_{1}=0, \label{eq:growth_einstein}
\end{align}
where we have defined for convenience:
\begin{align}
&\mu = 1+\frac{2 \beta_K^2}{K_X} \label{eq:kmou_mu}.
\end{align}
As a reminder, the subscript $\ttau$ denotes a derivative with respect to conformal time and $\mathcal{H}_E$ the conformal Hubble rate in the Einstein frame. $\mu$ is the modification to the cosmological Poisson equation from linear perturbations~\cite{brax_K-mouflage_2014_LSS},
\begin{equation}\label{eq:kmou_poisson}
    \nabla^2 \Psi_{\textrm N} = 4\pi G_{\textrm N} A(\phi) \mu \,a_E^2 \delta\rho,
\end{equation}
where $\Psi_{\textrm N}$ is the standard Newtonian potential. 
We can observe that in this frame, the scalar field provides additional contributions to the friction term in the growth equation, encapsulating the geodesic modifications to matter trajectories. Additionally, the deviation in the expansion history from \lcdm \ will also have an impact on the growth 
solution for K-mouflage\footnote{One can expose the Hubble factors by changing variables from conformal time to coordinate time.}. 

We can therefore tally 3 effects on the linear growth in K-mouflage theories that can cause deviations from \lcdm:

\begin{enumerate}
    \item The expansion history, $H_E$,
    \item The modification to the Poisson equation from linear perturbations in eq.~\eqref{eq:kmou_poisson}, $\mu$, 
    \item The effects of the conformal term,  $A$.
\end{enumerate}
These effects are summarised in the left panel of Fig.~\ref{fig:growth_breakdown}, where one can see that in K-mouflage gravity there is an overall enhancement in large-scale clustering compared to GR-\lcdm.

\subsection{Fifth force}\label{sec:fifth_force_einstein}

Similar to models with Vainshtein screening, K-mouflage models possess a derivative-type screening mechanism whereby the coupling of the scalar field to matter is effectively suppressed \cite{kobayashi_horndeski_2019}. The origin of this screening is the first derivative of $\phi$ in $K$ [defined in eq.~\eqref{eq:kmou_K}], which makes it differ from the Vainshtein mechanisms of DGP and the Cubic Galileon, where it is sourced by second derivatives of the scalar field \cite{joyce_beyond_2015}. The screening mechanism can be seen by considering the effective fifth force induced by the scalar field on a test particle orbiting a static, spherically symmetric mass distribution \cite{winther_fast_2015}. This results in

\begin{equation}
\label{eq:kmou_force}
    F_{\phi}(r) = \frac{G_N M(<r)}{r^2} \cdot \frac{2\beta_K^2}{K_X(r)},
\end{equation}
where $\beta_K$ was defined in eq.~\eqref{eq:conformal_factor} and is a constant.
$M(<r$) is the mass enclosed up to some radial distance $r$. K-mouflage screening therefore occurs when $K_X \gg 1$, suppressing the fifth force in eq.~\eqref{eq:kmou_force}. A parametric formalism for computing the forces in K-mouflage is detailed in \cite{lombriser_parametrisation_2016}, and its implementation in Einstein frame-based codes is discussed in~\cite{Bose:2024qbw}.

A discussion of the effect of the K-mouflage fifth force on clustering is presented in Section~\ref{sec:spectra_breakdown}.

\subsection{Constraints from Small Scales} 
\label{sec:constraints}

Having discussed some of the deviations of K-mouflage theories from GR-\lcdm{} (in the Einstein frame), we will now take the opportunity to review the status of K-mouflage with respect to existing observational tests of gravity, and consider restrictions on parameters on theoretical grounds.

In \cite{Barreira:2015_kmou_solar} it was shown that observations from the Cassini spacecraft and Lunar Laser Ranging (LLR) experiments could place strong bounds on the maximum magnitude of the fifth force. That is, these translate to bounds on the $2\beta_K^2/K_X$ term in eq.~\eqref{eq:kmou_force}. To satisfy LLR observations, which constrain more strongly than Cassini observations, $\beta_K \lesssim 0.1$. On even smaller scales, laboratory observations such as those from atom interferometer experiments constrain deviations from the Newtonian force to be smaller than $10^{-4}$~\cite{Sorrentino:2013uza}. Though it should be noted that on such scales, the effects of the fifth force are expected to be screened. K-mouflage theories also introduce a time-varying gravitational constant [this can be seen transparently in the Jordan frame, see eq.\eqref{eq:fifth_force}]. Thus, another observational constraint comes from the comparison of the gravitational constant at the time of Big Bang Nucleosynthesis (BBN) to that of today. Such an analysis implies $\beta_K \lesssim 0.22$~\cite{Barreira:2015_kmou_solar}. In light of these constraints, we focus on $\beta_K \in \{0.1,0.2\}$. We keep $\beta_K = 0.2$, despite the implications of LLR, in order to maximally display the phenomenological effects of K-mouflage.

Stability considerations can also play a role in restricting the freedom of the model parameters. In order to avoid ghost instabilities, one requires that $K_X > 0$~\cite{Brax:2015pka}. For this reason, in this paper we focus on, positive values for the coefficient of the term with the highest order in eq.~\eqref{eq:kmou_K}, i.e. $K_0>0$. We also require that $A > 0$, to avoid instabilities associated with $A$ possibly crossing zero. This is satisfied by the exponential choice for $A$ defined by eq.~\eqref{eq:conformal_factor}. Finally, it was noted in \cite{brax_K-mouflage_2014_background} that the scalar field energy density is subdominant at early times if $K$ has a power law relationship with $X$. Therefore, to restrict the cosmological effects of the models we study in this paper to late time, we only consider $n \in \mathbb{N}, n>0$.


\section{K-mouflage in the Jordan frame}
\label{sec:kmou_Jordan}

As outlined in Section~\ref{sec:HiCOLA}, \HiCOLA is designed to work with members of the reduced Horndeski class, which have actions of the form of eq.~\eqref{eq:rH_action}. While this action nominally includes the K-mouflage subclass, a difference between the Einstein frame action of eq.~\eqref{eq:kmou_action} and the Jordan frame action of eq.~\eqref{eq:rH_action} is that in the latter, matter is \textit{minimally} coupled to the metric. There is now also a conformal term, $G_4$ multiplying the Einstein-Hilbert term. Thus, implementing K-mouflage in \HiCOLA requires establishing the form of K-mouflage in the Jordan frame.

In the Jordan frame, the Horndeski functions of K-mouflage theories are~\cite{benevento_K-mouflage_2019}:

\begin{align}
\label{eq:kmou_J}
    K_J &= \frac{M_K^4}{A^4}K + \frac{6M_K^4 \mplanck^2}{A^5} X \left( \frac{2}{A} \left( \frac{dA}{d\phi}\right)^2 - \frac{d^2A}{d\phi ^2} \right),\\
    G_3 &= -\frac{3 \mplanck^2}{A^3} \left( \frac{dA}{d\phi} \right)\label{eq:kmou_J_G3},\\
    G_4 &= \frac{\mplanck^2}{2A^2}.
\end{align}
We can see that in the Jordan frame, K-mouflage theories have a non-zero $G_3$ term, in principle allowing for the influence of second derivative effects, which might be considered `Vainshtein-like' behaviour. However, we will see in Section~\ref{sec:kmou_jordan_force} that we do not find Vainshtein contributions to the screening factor because $G_{3X}\equiv \partial_X G_3 = 0$. Additionally, unlike the archetypal shift-symmetric theories studied in \cite{Wright:2022krq} -- in which the Horndeski functions only depend on $X$ -- in K-mouflage theories they explicitly depend on $\phi$. There is also a non-trivial conformal term, $G_4(\phi)$, which was held constant in models we previously studied \cite{Wright:2022krq}. In order to understand K-mouflage's unique clustering properties, the Cubic Galileon will be used as a point of comparison when discussing results in Section~\ref{sec:power_spectra}.

In this section, we will focus on the K-mouflage results with $n=2$, $K_0 =1$ and $\beta_K=0.2$ as it displays the strongest modified gravity effects, for ease of presentation. We choose to tune $\lambda$ in the Einstein frame so that our results can be compared with those in \cite{hernandez-aguayo_fast_2022}. But one could do this in the Jordan frame instead, and \HiCOLA can support either approach. We also choose to evaluate GR-\lcdm{} results at Einstein redshifts for comparability with existing results.

\subsection{Einstein and Jordan frames}

With coordinate invariance as a guiding principle of GR, transformation to convenient frames for particular computations has long been exploited. Two well-known frames in the study of scalar-tensor gravity are the Jordan and Einstein frames, whose relation is mediated by the conformal factor, $A(\phi)$:

\begin{equation}
\label{eq:E2J_metric_relation}
    g_{\mu \nu} = A^2(\phi) \tilde{g}_{\mu \nu}.
\end{equation}
As a reminder, $g_{\mu \nu}$ (without the tilde) is the Jordan frame metric. In the Einstein frame, one recovers the standard Einstein field equations, and therefore in the context of numerical simulations, one would find the standard Poisson equation relating the gravitational potential to matter density. However, matter would not follow standard geodesics, but instead follow paths modified by the conformal factor $A$ \cite{koyama_modified_2019}. Conversely, in the Jordan frame the field equations may differ substantially from the Einstein field equations of GR; however, matter remains minimally coupled to the metric.

These frame differences are unphysical and therefore cannot affect any observables. However, one will encounter differences for any intermediate quantities, such as the Hubble rate. A summary of these differences is presented here, though we refer readers to \cite{francfort_cosmological_2019} for a detailed discussion.

\subsection{Background}

Eq.~\eqref{eq:E2J_metric_relation} implies, for an FLRW metric, that the scale factor in each frame is related by:   
\begin{equation}
\label{eq:E2J_scalefactor}
    a_J = A\left( \phi\left( a_E \right) \right) \cdot a_E.
\end{equation}
This indicates that the scale factor, or redshift, is an ambiguous quantity when comparing results between the Einstein and Jordan frames. Care must be taken to ensure that results are being compared fairly lest one, for instance, interpret pathological relative enhancement or suppression of structure that is actually coming from a mismatch in the times being compared. It is worth remembering that as this transformation depends on the background solution for the scalar field, it is dependent on the parameter values of the given K-mouflage model being studied. 

This has a number of practical consequences when studying K-mouflage in the Jordan frame. While the values of the scalar field remain identical between the frames, they are assigned to different times. This therefore affects any derivatives and also any numerical interpolations performed. We distinguish between these cases by the notation {$\phi_J: \{ a_J \} \rightarrow \{\phi\}$}  and {$\phi_E: \{a_E\} \rightarrow \{\phi\}$}.

In a similar vein, the relationship between the Einstein and Jordan values for the Hubble rate is give by \cite{francfort_cosmological_2019}: 
\begin{equation}
\label{eq:E2J_hubble}
    H_{\rm J}(a_E) = \frac{H_{\rm E}(a_E) }{A(\phi(a_E))} \left[ 1 + \frac{\beta_{\rm K}}{\mplanck} \phi'_E(a_E) \right] \, . 
\end{equation}
The scale-factor dependence has been shown explicitly. We remind the reader that an extra step is needed to express the Hubble rate with respect \textit{Jordan} scale factor values, $H_J(a_E) \rightarrow H_J(a_J)$.

\begin{figure}
    \centering
    \centering
    \makebox[\textwidth][c]{
    \includegraphics[width=1.05\textwidth]{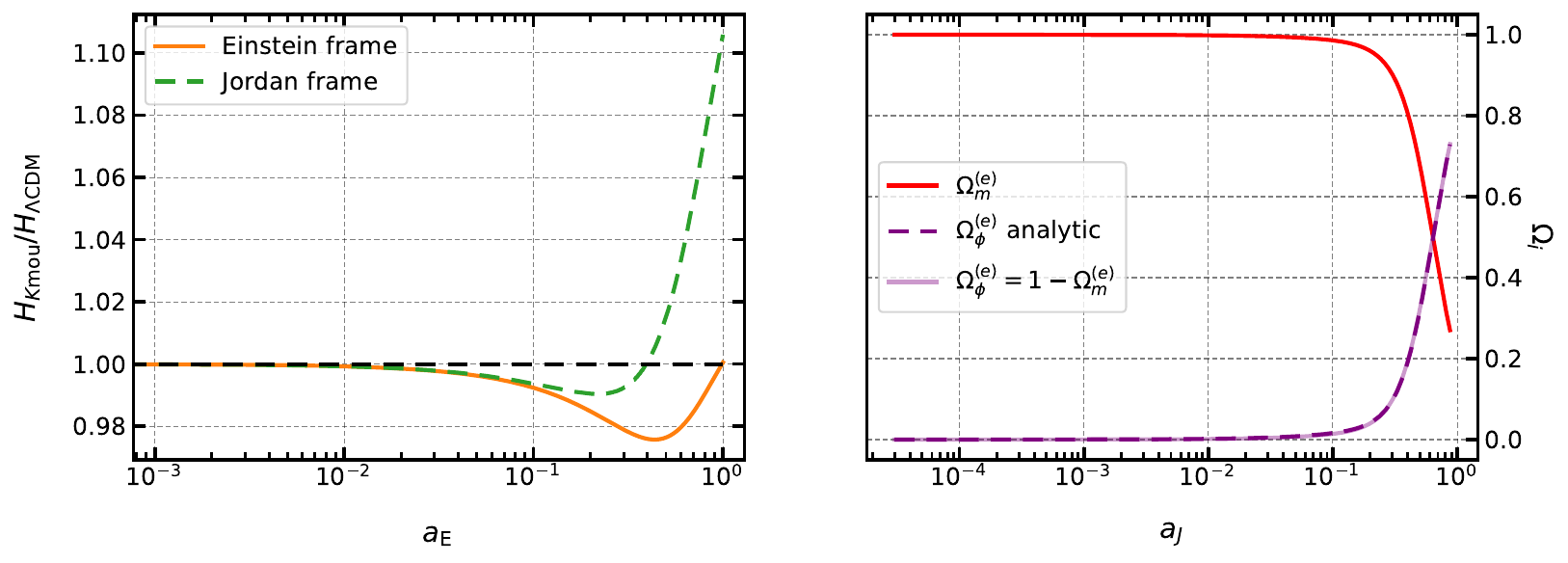}}
    \caption{\textit{Left panel}: Ratio of the Hubble expansion rates in the Einstein (solid orange line) and Jordan (dashed green line) frames for K-mouflage gravity with the the Hubble rate in $\Lambda$CDM. In this case, the K-mouflage parameters are: $n=2, K_0=1$ and $\beta_K = 0.2$.\\
    \textit{Right panel}: Fractional energy densities for K-mouflage in the Jordan frame, for the same model parameters as in the left panel. Notice that it is the \textit{effective} energy densities, defined in eqs.~\eqref{eq:effective_omegam} and \eqref{eq:effective_omegaphi}, that sum to 1 in the Jordan frame. The need to define effective energy densities to recover regular closure is due to the transformation of energy densities between the Jordan and Einstein frame.}
    \label{fig:kmou_hubble_omega}
\end{figure}

For the K-mouflage model we consider in this work, the translation of the Hubble rate between frames translates to the map shown in the left panel of Fig. \ref{fig:kmou_hubble_omega}.  We highlight two key features of this plot, which shows the ratio of the K-mouflage Hubble rate to that of \lcdm. Firstly, there is a shift in the trough of the curve between frames, i.e. the scale factor at which maximum deviation from \lcdm{} expansion differs. Secondly, in the Jordan frame, the ratio rises above 1, i.e. the target value of $H_0$ is not recovered (recall we tuned the parameter $\lambda$ to recover a target $H_0$ in the Einstein frame). We deduce that the effects of the background expansion history on the formation of LSS can differ between Einstein and Jordan frames. From $a_E \sim 0.4$, the Jordan frame sees relative \textit{suppressive} effects from the faster-than-\lcdm{} expansion history, in contrast to the growth-enhancing effects from slower expansion in the Einstein frame.

Since the energy densities transform between the Einstein and Jordan frame, the fractional energy densities no longer sum to 1, as implied by eq.~\eqref{eq:kmou_friedmann} in the Einstein frame. Instead, the Friedmann-analogue in the Jordan frame (expressed as a closure equation) is~\cite{benevento_K-mouflage_2019}\footnote{Where $\epsilon$ in this paper is $\epsilon_2$ in \cite{benevento_K-mouflage_2019}.}:
\begin{align}
    1 &= \frac{A^2}{(1-\epsilon)^2}\bigg( \Omega_{\rm m}(a) + \Omega_r(a) + \Omega_{\phi}(a) \bigg),\\
    \epsilon &= \frac{d \ln A}{ d (\ln a_J)},
\end{align}
where eq.~\eqref{eq:kmou_rhophi} is used to define the ``analytic'' expression for $\Omega_{\phi}$ featured in the right panel of Fig.~\ref{fig:kmou_hubble_omega}:
\begin{equation}
    \Omega_{\phi} = -\frac{\lambda^2 K}{3E^2}  + \frac{\phi^{'2} K_X}{3}.
\end{equation}
We can therefore define effective fractional energy densities which do sum to 1, defined in eqs.~\eqref{eq:effective_omegam} and~\eqref{eq:effective_omegaphi}: 

\begin{align}
    \Omega^{(e)}_{\rm m} &= \frac{A^2}{(1-\epsilon)^2} \Omega_{\rm m},\label{eq:effective_omegam}\\
    \Omega^{(e)}_{\phi} &= \frac{A^2}{(1-\epsilon)^2} \Omega_{\phi}.\label{eq:effective_omegaphi}
\end{align}

As can be seen in the right panel of Fig.~\ref{fig:kmou_hubble_omega}, the light purple curve and the dashed dark purple curve are coincident, verifying the satisfaction of closure in the familiar sense.

The frame transformations, however, do not affect the rising significance of the scalar field at late times, thereby facilitating a dark energy era today.

\subsection{Linear Theory}\label{sec:linear_theory}

Having understood the importance of the frame transformation for background expansion quantities, we will now study its effects on linear growth.
The growth equation for the K-mouflage model in the Jordan frame reads:

\begin{align}
&D_{1,\tau \tau} + \mathcal{H}_J D_{1,\tau} - \frac{3}{2} \, \frac{G_{G4}}{G_N} \Omega_{\rm m}(a_J) H_{0}^{2} a_J^{2} \mu D_{1} = 0.\label{eq:growth_jordan}
\end{align}
Note that $\tau$ is the conformal time in the Jordan frame. The scalar field contributions in the Jordan frame now appear in terms that originate from the Poisson equation multiplying $D_1$, $\mu$. $\Omega_{\textrm{m}}$ is the Jordan frame matter density; the presence of $G_{G4}/G_N = A^2$ multiplying it can be interpreted as an effective transformation of the Einstein-frame matter density in eq.~\eqref{eq:growth_einstein} to the Jordan-frame matter density.

Comparing eqs.~\eqref{eq:growth_einstein}, \eqref{eq:E2J_hubble}, and \eqref{eq:growth_jordan} we can summarise that we expect \textit{frame} differences to  come from:
\begin{enumerate}
    \item The Hubble rate transformation, $H_J$ vs. $H_E$,
    \item The modifications to the coefficients of $D_{1,\tau}$ and $D_1$ [compare eq.~\eqref{eq:growth_einstein} and eq.~\eqref{eq:growth_jordan}] due to the conformal term. 
\end{enumerate}

\begin{figure}
    \centering
    \makebox[\textwidth][c]{\includegraphics[width=1.1\textwidth]{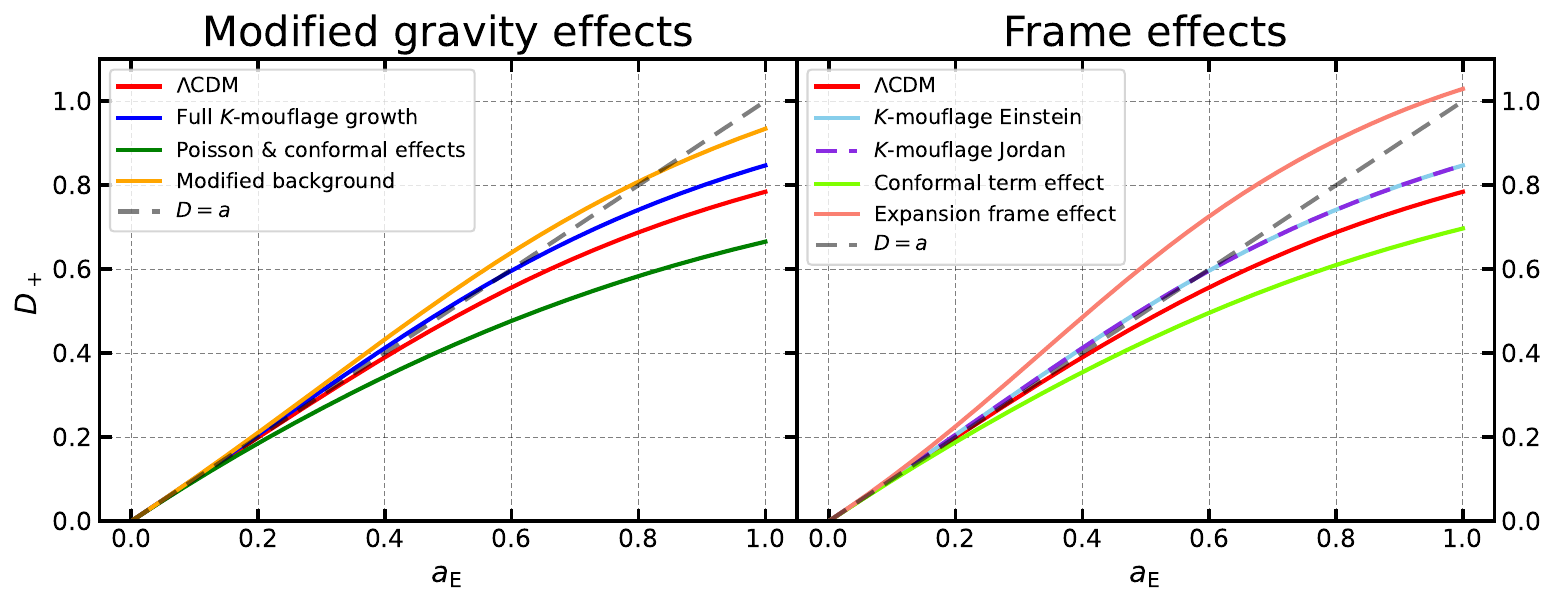}}
    \caption{\textit{Left:} Growing mode solutions, $D_+$, of the growth equation for K-mouflage in the Einstein frame (blue), and GR-$\Lambda$CDM (red). The modifications to the K-mouflage growth factor equation from the coupling term in the Poisson equation and the conformal term  are shown in green,  while the enhancing effects from the expansion history are shown in yellow. 
    The non-linear counterpart of this plot is Fig.~\ref{fig:bk_kmou_breakdown_evolving_redshift}. \\ \textit{Right:} The opposing contributions of the conformal term in the K-mouflage growth factor equation in the Jordan frame [eq.~\eqref{eq:growth_jordan}] and Einstein frame [eq.~\eqref{eq:growth_einstein}], and the transformation of the expansion history [eq.~\eqref{eq:E2J_hubble}], resulting in negligible frame differences.
    }
    \label{fig:growth_breakdown}
\end{figure}

The frame differences are depicted in the right panel of Fig.~\ref{fig:growth_breakdown}. We can see an appreciable enhancement in linear growth for K-mouflage theories relative to \lcdm \ (blue/purple-dashed curve vs. red curve, respectively). We can also see that there are negligible frame differences in this result. Given the significantly different Hubble rates between frames, this implies that the conformal contributions to the growth compensate against this difference, which we can see through the opposite placements of the orange and green curves in the right panel of Fig.~\ref{fig:growth_breakdown}. The similarity of the growth between frames is expected, as on sub-horizon scales the power spectrum is approximately independent of conformal transformations~\cite{francfort_cosmological_2019}.

\begin{figure}
    \centering
    \includegraphics[width=0.8\textwidth]{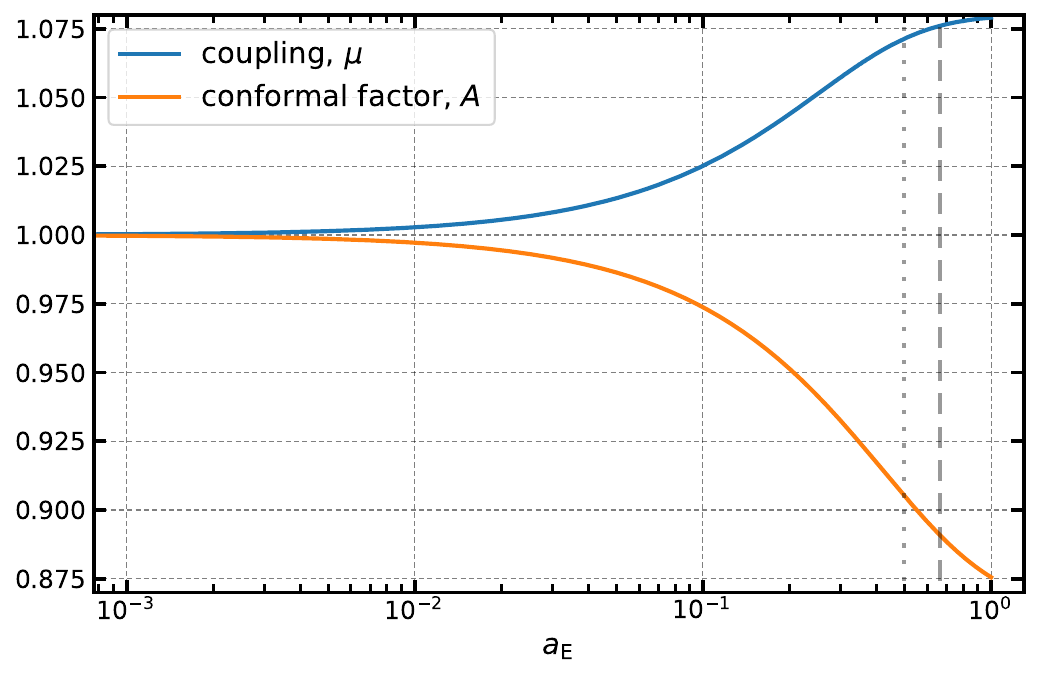}
    \caption{Components affecting the growth equation, besides the expansion: the conformal factor $A$ [eq.\eqref{eq:conformal_factor}], and the modification to the Poisson equation $\mu$ [eq.~\eqref{eq:kmou_mu}]. The dashed vertical line corresponds to a redshift of $z_E = 0.5$, while the dotted vertical line corresponds to a redshift of $z_E = 1.0$. These will be referred to in the discussion of Fig.~\ref{fig:bk_kmou_breakdown_evolving_redshift}.}
    \label{fig:linear_effects}
\end{figure}

The modified gravity effects of K-mouflage with respect to GR-\lcdm{} are shown in the left panel of Fig.~\ref{fig:growth_breakdown}. In green is the combined effect of the coupling $\mu$ and the conformal term $A$ on the growth. The conformal factor, being a function that decreases with time (as in Fig.~\ref{fig:linear_effects}), acts to weaken clustering by weakening the effective gravitational constant. It should be noted that this not only affects any fifth force contributions, but also the Newtonian forces experienced by particles. Meanwhile, we can see that the coupling $\mu$ increases with time above $1$, indicating that linear perturbations act to strictly enhance large-scale clustering. Overall, there is a suppression to growth, and this is largely because the conformal term's suppression majorly outweighs the small enhancements from the coupling. Meanwhile, in yellow is the effect of the deviations in the K-mouflage expansion history relative to \lcdm. We see that the expansion history enhances the growth, and together with the enhancement from the coupling, this is enough to compensate for the suppression from the conformal factor. Thus, there is net enhancement in the K-mouflage growth compared to GR-\lcdm.

Whilst in this work we focus on models with enhanced growth compared to GR-$\Lambda$CDM, gravity theories that have suppressed formation of structures at late times can be used to alleviate the $S_8$ tension~\cite{Lin:2023_S8tension_suppressed_growth}. See \cite{benevento_K-mouflage_2019} for a discussion on the K-mimic model, which maintains the suppressive effects sourced by the conformal factor whilst minimising the enhancement coming from the altered expansion history compared to $\Lambda$CDM. Also see \cite{cataneo:2024_mochiclass_suppressed_growth} for other examples from the Horndeski class found parametrically.

\subsection{The K-mouflage force}
\label{sec:kmou_jordan_force}

In addition to the modified expansion history and linear modifications to the Poisson equation, we also need to consider modifications to the forces experienced by the dark matter particles in the $N$-body part of the simulation. The force implementation in \HiCOLA was described in general terms in section \ref{sec:HiCOLA}. Here we will specialise to the K-mouflage case, and for convenience we reproduce the schematic representation of the force (see \cite{Wright:2022krq} for the definitions in terms of the Horndeski functions): 
\begin{equation}
\label{eq:fifth_force_gg4}
    F_{\textrm{tot}} = F_N \frac{G_{G4}}{G_N}\left(1+\beta S\right),
\end{equation}
where
\begin{equation}
    G_{G4} = (16\pi G_4)^{-1}.
\end{equation}
Recall that the coupling $\beta$ is the linear enhancement to the regular Newtonian force arising from the scalar field; $S$ is the screening factor, which recovers unscreened environments when $S\rightarrow 1$ and fully screened environments where $S\rightarrow 0$. $G_{G4}/G_N$ acts as an effective rescaling of Newton's gravitational constant and is sourced by $G_4$ in eq.~\eqref{eq:rH_action}.

The definitions given in \cite{Wright:2022krq} were derived specifically in the context of Vainshtein screening, and so one may expect that the screening factor $S$ must be re-derived for K-mouflage, much like the coupling $\beta$. However, an alternative answer lies in a change in perspective. When transformed into the Jordan frame as per eq.~\eqref{eq:kmou_J}, we see that the K-mouflage model we study has a non-zero $G_3$ term, nominally implying it could contain Vainshtein screening. Furthermore, in Section~\ref{sec:Kmou_vs_Vainshtein}, we will see that in relative terms, the K-mouflage mechanism always manifests on scales even smaller than those of the Vainshtein mechanism, and therefore arguably beyond the scales that \HiCOLA can probe. It is therefore sufficient to consider the Vainshtein definition for $S$, using the Horndeski forms defined in eq.~\eqref{eq:kmou_J}.  Perhaps unsurprisingly, this calculation gives $S \rightarrow 1$, since $G_{3X}=0$. The explanations for this are in section~\ref{sec:Kmou_vs_Vainshtein} [also see eq.~\eqref{eq:screen_coeff}]. This is therefore consistent with the Einstein frame calculations of \cite{brax_K-mouflage_2014_LSS}, where it was shown that quasi-linear objects are unscreened in K-mouflage models.

The novel aspect of the K-mouflage model we study that sets it apart from  Vainshtein models in \cite{Wright:2022krq}, aside from its differing expansion history and linear modification, is its conformal effect on the effective gravitational constant. $G_{G4}/G_N$ is no longer trivially $1$, and is instead the square of the time-dependent conformal function, $A$, plotted in Fig.~\ref{fig:linear_effects}. This translates into a modification of even the \textit{Newtonian} forces experienced by particles, which persists on all scales.

\section{K-mouflage vs. Vainshtein Screening Scales}
\label{sec:Kmou_vs_Vainshtein}

We saw in eq.~\eqref{eq:kmou_J_G3} that the Jordan-frame Horndeski functions for K-mouflage feature a non-zero $G_3$ term, which is normally associated to Vainshtein-like behaviour. However, we have not yet shown that K-mouflage in the Jordan frame can be treated inside \HiCOLA{} in the same manner as Vainshtein theories \cite{Wright:2022krq}. In this section, we will summarise why a generic fifth force law cannot be derived for K-mouflage theories, and why our implementation must instead be limited to a chosen functional form for $K$ in eq.~\eqref{eq:kmou_action}. But, we will also see that K-mouflage screening takes place on scales smaller than Vainshtein screening scales, implying that it is sufficient to treat K-mouflage as a theory with a Vainshtein screening factor, at least until very small scales which are beyond the accuracy limit of \HiCOLA. We refer readers to Appendix~\ref{sec:Kmou_vs_Vainshtein_appendix} for the details behind this discussion.

For readers that wish to directly see the results of this approach on the non-linear matter power spectra, see Section~\ref{sec:power_spectra}.

We begin by considering a generalised Galileon-type action, where we write down patterned sequences of terms

\begin{align}
\label{eq:galileon_action}
    S = \int d^4x \sqrt{-g}\left( M_P^2 R + \right. &\left[ X + \frac{X^2}{\Lambda_2^4} + \frac{X^3}{\Lambda_2^8} + ... + \frac{X^m}{\Lambda_2^{4m-4}} + ... \right] \\
    +&\left[ \frac{X\Box \phi}{\Lambda_3^3} + \frac{X^2\Box \phi}{\Lambda^4_2 \Lambda_3^3} + ... +\frac{X^n\Box \phi}{\Lambda_2^{4n-4}\Lambda_3^3} + ... \right] \nonumber\\
     +& \left[ \frac{X(\Box \phi)^2}{\Lambda_3^6} + \frac{X(\Box \phi)^3}{\Lambda_3^9} + ... + \frac{X(\Box \phi)^p}{\Lambda_3^{3p}} + ...\right]  \nonumber\\
     +&\left. \left[\frac{X^2(\Box \phi)^2}{\Lambda^4_2 \Lambda_3^6} + \frac{X^3(\Box \phi)^2}{\Lambda_2^8 \Lambda_3^6} + ... +\frac{X^q(\Box \phi)^2}{\Lambda_2^{4q-4}\Lambda_3^6} + ... \right] + ...\right). \nonumber
\end{align}
$\Lambda_2$ is the mass scale associated with $X$, and $\Lambda_3$ is the mass scale associated with $\Box \phi$. As we are interested in cosmological scales, derivatives are of order $H_0$. Hence, the Ricci scalar, composed of time derivatives for an FLRW metric, is of order $H_0^2$. For the Ricci term and the first term in eq.~\eqref{eq:galileon_action} to be of comparable order, we must have
$M_P^2 H_0^2 \sim \Lambda_2^4$. On similar grounds, we can reason that $\Lambda_3^3 \sim M_P H_0^2$.

By considering a perturbation of the scalar field in eq.~\eqref{eq:galileon_action} given a perturbed FLRW metric in the Newtonian gauge, we can establish on what distance scales terms in the perturbed action are comparable to each other, using the equation of motion for the scalar field perturbation. In particular, we will be most interested in comparing the high-order perturbations to the quadratic (in $\phi$) perturbation sourced by $X$.

\underline{From the $X^m/\Lambda^{4m-4}$} (`K-mouflage') set of terms we find the distance scale where the $m=2$ is comparable to the $m=1$ term:
\begin{equation}
\label{eq:rX2}
     r_{X^2} \coloneqq \sqrt{\frac{r_g}{H_0} }.
\end{equation}
where
\begin{equation}
    r_g = G_N M.
\end{equation}
$M$ is the mass of the source giving rise to the spacetime curvature and scalar field profile. In other words, the perturbation sourced by $X^2$ is as significant as the perturbation from the quadratic term $X$ on distance scales less than $r_{X^2}$. In fact, it can be shown that eq.~\eqref{eq:rX2} is true for all $m \in \mathbb{N}$. This means all terms $\{X^{m}\}$ in eq.~\eqref{eq:galileon_action} are as significant as the quadratic term on the same scales. 

\begin{figure}
    \centering
    \includegraphics[width=\textwidth]{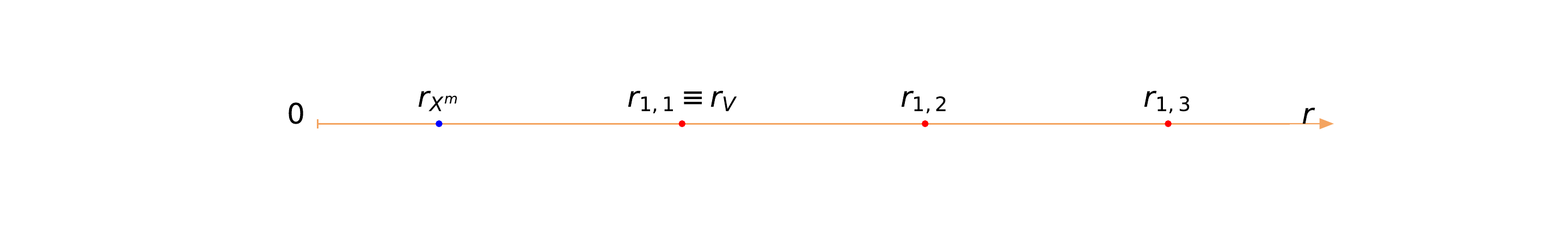}
    \caption{Diagrammatic representation of the radii of significance for $X^m$ terms compared to $X^p(\Box \phi)^q$ terms. $r_{X^m}$ is the radius of significance for the perturbation sourced by all terms like $X^m$ in action~\eqref{eq:galileon_action}. $r_{p,q}$ corresponds to the radius of significance for terms like $X^p(\Box \phi)^q$. The $p=q=1$ term is the canonical Vainshtein radius $r_V$. Notice the $X^p(\Box \phi)^q$ terms have differing radii of significance, in contrast to the $X^m$ terms.}
    \label{fig:numberline}
\end{figure}

\underline{From the $X (\Box \phi)^q / \Lambda_3^{3q}$} (`Vainshtein') set of terms we find:
\begin{equation}
    r_V \coloneqq \left( \frac{2 r_g}{H_0^2} \right)^{1/3} > r_{X^m}.
\end{equation}
This implies there is a region, $r_{X^2} < r < r_V$, where  the perturbations sourced by $X\Box\phi$ are comparable to the quadratic term, but those from $\{X^m\}$ are not and can be neglected. Furthermore, a hierarchy is present for terms of the form $X(\Box \phi)^q$ for increasing $q$. Fig.~\ref{fig:numberline} summarises the results of these comparisons of the perturbations.

Thus, from these calculations we can take away two major results. The first is that one \textit{cannot} perform the same quasilinear perturbative calculation done in \cite{Kimura:2011dc,Wright:2022krq} to derive an analogue of the generic Vainshtein fifth force for K-mouflage theories. There is no hierarchy with which one can choose to truncate to a certain order to include the major sources of screening on small scales. In K-mouflage theories, the force derived in such a fashion will depend on the functional form of $K$, and change with it. The second is that there exists a range of scales, $r_{X^m} < r < r_V$ where the $X \Box \phi / \Lambda_3^3$ term is the driver of screening behaviour and dominates over behaviours sourced by $\{X^m\}$. 

By comparing $r_V$ and $r_{X^m}$, it can be deduced that the ratio of Vainshtein radius to the `K-mouflage' radius is of the order of the Hubble radius divided by the Schwarzschild radius for a given object. For galactic clusters, the latter is much smaller, and therefore $r_{X^m}$ is \textit{several} orders of magnitude smaller than $r_{V}$. In \cite{Wright:2022krq}, the Vainshtein radius for Cubic Galileons was observed to be at roughly $1.4$ $h^{-1}$Mpc; meanwhile \HiCOLA{} was observed to be accurate up until roughly $0.8$ $h^{-1}$Mpc. The K-mouflage screening scales can therefore be reasoned to be beyond the scales that \HiCOLA{} can accurately probe.
It is therefore sufficient to treat K-mouflage theories by considering only their behaviour in this intermediate regime where Vainshtein behaviour dominates.

\section{\HiCOLA \ Results for Non-linear Scales}
\label{sec:power_spectra}

\subsection{Simulation settings}
For the results in this paper, particles were initialised at $z=19$ by applying a \lcdm{} back-scaling to a $z=0$ linear power spectrum computed with \texttt{class} \cite{blas_cosmic_2011}. To include high-redshift effects of K-mouflage gravity in the initial conditions, the backscaled \lcdm{} power spectra were re-scaled with the factor $(D_1^{\rm Kmou}(z=19)/D_1^{\rm \Lambda CDM }(z=19))^2$.

The cosmological parameters for this input spectrum and also the background solution data are summarised in Table~\ref{tab:base_cosmo_params}, matching those used in \cite{hernandez-aguayo_fast_2022}. As we consider ratios of power spectra for the non-linear clustering results in this paper and only focus on late-time effects, we neglect the effects of radiation in the K-mouflage power spectrum. This does not affect power spectra ratios provided the same is also done for the GR-\lcdm{} spectra, which is the case in this paper. 

The key \HiCOLA \ simulation settings were as follows:
\begin{itemize}
    \item Number of particles ($N_{\textrm{part}}$): $512^3$,
    \item Box size: $400$ Mpc/h,
    \item Number of time steps: $40$ (see Sec.~\ref{sec:convergence} for a discussion on this choice),
    \item Force mesh size: $3(N_{\textrm{part}})^{1/3} = 1536$. We make this choice following the analysis in \cite{izard_ice-cola_2016} where this was found to be optimum for COLA simulations for balancing accuracy and memory demands.
\end{itemize}

\begin{table}
\begin{center}
\begin{tabular}{ll}
\toprule
$\Omega_{\rm m0}$   &  0.3089    \\ 
$\Omega_{\rm b0}$ & 0.0486 \\ 
$h$ & 0.6774 \\ 
$n_s$ & 0.9667 \\ 
$p_s$ & $0.05 \, {\rm Mpc}^{-1}$\\ 
$A_s$ & $2.065 \cdot 10^{-9}$ \\ 
\bottomrule
\end{tabular}
\caption[Base cosmological parameters]{Base cosmological parameters used throughout, matching those used in \cite{hernandez-aguayo_fast_2022}. The effects of radiation are effectively excluded in the computations of the background and clustering in this paper. As we focus on late-time effects in this paper, this does not significantly affect the conclusions drawn on clustering behaviour.}
\label{tab:base_cosmo_params}
\end{center}
\end{table}

\subsection{Main results and phenomenology}
\label{sec:spectra_breakdown}

\subsubsection{Boost definition}
We are now in a position to assess the matter power spectra for K-mouflage, depicted in the left panel of Fig.~\ref{fig:bk_breakdown} as a ratio with respect to a GR-$\Lambda$CDM power spectrum, i.e. a boost. As a reminder, as we produce results for K-mouflage in the Jordan frame, the boost ratios we analyse are of the form,

\begin{equation}\label{eq:boost}
    B(k) = \frac{P_{\textrm{kmou}}(z_J)}{P_{\textrm{GR}}(z_E)}.
\end{equation}

We choose to evaluate the GR-\lcdm{} power spectra at Einstein redshifts instead of Jordan for the practical reason of being able to compare \HiCOLA{} results with those in \cite{hernandez-aguayo_fast_2022}. This is simply a convention, and one could equally evaluate both spectra in eq.~\eqref{eq:boost} in the Jordan frame, as in \cite{Brax:2015lra}. One could in principle avoid this ambiguity altogether by working only with frame-independent observables.

The relationship between the Jordan and Einstein redshifts follows from eq.~\eqref{eq:E2J_scalefactor}:

\begin{equation}
    z_J = \frac{1+z_E}{A} - 1.
\end{equation}

\subsubsection{Contributions of modified background and forces}
\begin{figure}
    \centering
    \makebox[\textwidth][c]{\includegraphics[width=1.1\textwidth]{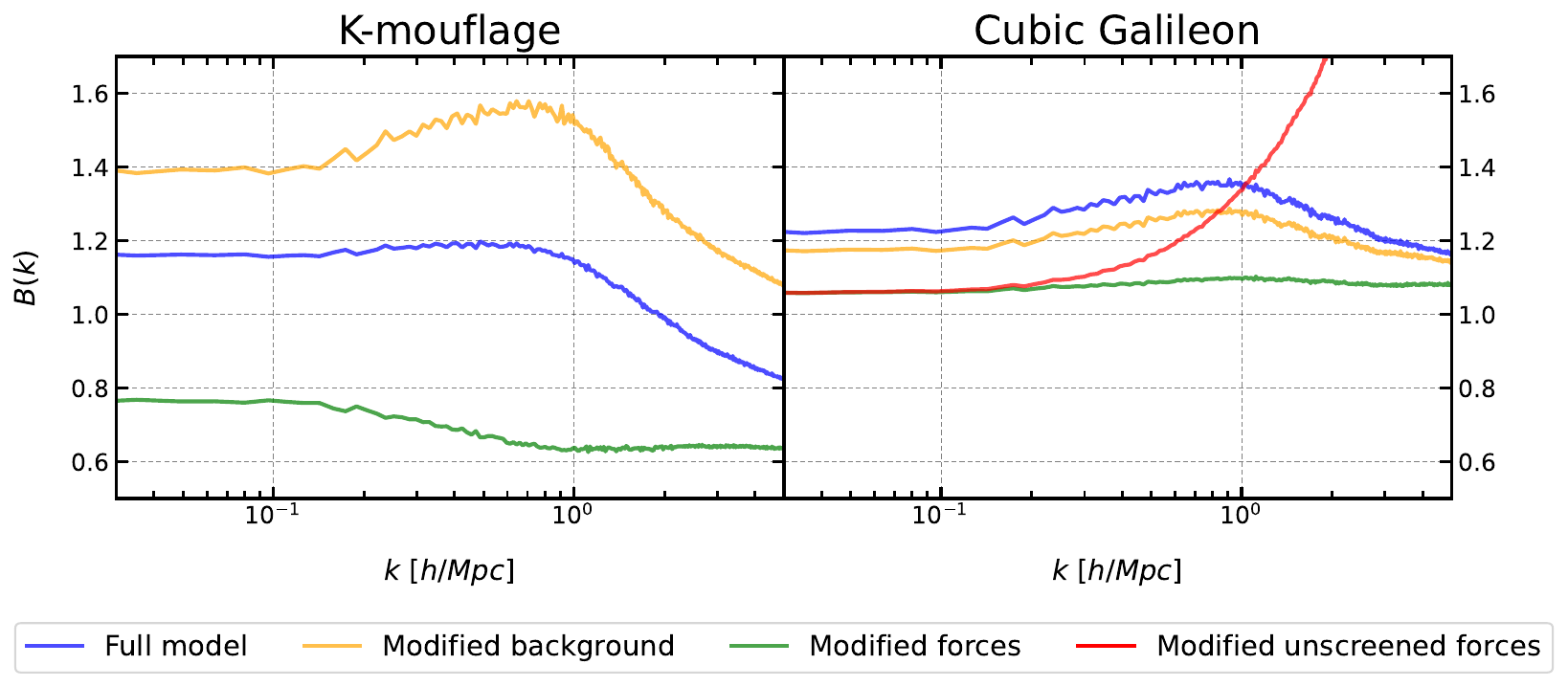}}
    \caption{A comparison of the effect of modifications to the background (yellow curves), and modifications to the gravitational forces (green curves) experienced by dark matter particles, on the deviations from GR-\lcdm \ at the level of the power spectrum. On the left is K-mouflage with $n=2$, $K_0=1$ and $\beta_K=0.2$ (the K-mouflage model with the strongest boost signal), and on the right the Cubic Galileon for reference. As the Cubic Galileon possesses Vainshtein screening, there is an extra curve (red), showing the effects of the \textit{unscreened} fifth force on the Cubic Galileon boost. For both models, the modified background is a significant contributor to the enhancement of the boost. However, unlike the Cubic Galileon, where the Vainshtein mechanism suppresses the enhancing effects of the linear fifth force, in K-mouflage the strong modified background-enhancement competes with the conformal weakening of the gravitational force. This leads to a suppression of the boost, below $1$ in K-mouflage's case.}
    \label{fig:bk_breakdown}
\end{figure}

In Fig.~\ref{fig:bk_breakdown} we categorise the K-mouflage modified gravity effects as those coming from the modified cosmological background (orange), and those coming from the modifications to the gravitational force (green). The full K-mouflage model corresponds to the curve with both effects included (blue). In this section we focus on the set of parameters that gave the strongest deviations from GR-\lcdm: $n=2$, $K_0=1$, $\beta_K=0.2$. For reference, a Cubic Galileon boost is also plotted in the right panel of Fig.~\ref{fig:bk_breakdown}, generated with the same cosmological parameters. The coefficients of $K$ and $G_3$ for the Cubic Galileon were $k_1 = -1$ and $g_{31}=0.1637$ respectively (see \cite{Wright:2022krq} for the definition of the Cubic Galileon and its parameters). These values lie in the tracker region of the Cubic Galileon parameter space (see Appendix B2 of \cite{Bose:2024qbw}); the tracker being convenient to use on account of the stability of the background equations of motion. This made it possible to easily match the cosmological parameter values used for the K-mouflage results.

We begin by first confirming what we expected: that on large scales, there is an overall scale-independent enhancement in K-mouflage with respect to GR-\lcdm. This can be seen through the blue and yellow curves in the left panel of Fig.~\ref{fig:bk_breakdown}, the full boost signal and contributions from the modified background respectively, lying horizontally above $1$ for low $k$. This is consistent with what we saw in Fig.~\ref{fig:growth_breakdown}; we find a $16.9\%$ enhancement on large scales in the boost, compared to $16.8\%$ as computed from the square of the ratio of growth factors in K-mouflage to GR-\lcdm\footnote{The small differences can be attributed to sampling variance effects when measuring the power spectrum from the \HiCOLA snapshots.}. 

Perhaps what is most fascinating at first glance about the full K-mouflage boost, the blue curve in the left panel of Fig.~\ref{fig:bk_breakdown}, is that on non-linear scales it does \textit{not} continually increase like the Cubic Galileon case without screening, i.e. like the red curve of the right panel. This is despite K-mouflage not possessing any screening effects on these scales, as discussed in Sections~\ref{sec:kmou_Jordan} and~\ref{sec:Kmou_vs_Vainshtein}. Of course, the vague similarity in the shape of the full K-mouflage boost and the full Cubic Galileon boost should not be taken to mean that they share screening effects. Indeed, we can see that the K-mouflage boost curve does not return to 1 on small scales, but falls below it. 

What we see is that the balance between the enhancing effects of the modified background and fifth force contributions, and the suppressive effects of the weakening of the gravitational constant (due to the conformal factor) has tipped in the other direction on small scales. On large scales, we see that the combined modified background and coupling effects are dominant, leading to enhancement and matching what we saw in the left panel of Fig.~\ref{fig:growth_breakdown}. However, on small scales we see that the modified background enhancement weakens, as shown by the yellow curve in the left panel of Fig.~\ref{fig:bk_breakdown}. This coincides with an increase in the suppression from the modified K-mouflage forces, shown by the green curve. We see that the boost with only modified forces drops from a large-scale suppression of approximately $75\%$ to approximately $65\%$ by $k \sim 1$ h/Mpc. 
This should be contrasted with what occurs for the Cubic Galileon. We can see in the right panel of Fig.~\ref{fig:bk_breakdown} that \textit{all} the modified gravity components contribute as enhancements. Here, the screening mechanism plays a vital role in tempering the contributions from the coupling term of the Cubic Galileon, and furthermore, ensuring that the force returns to the Newtonian force as would be found in GR. This is not what occurs for K-mouflage, as the conformal term also weakens the Newtonian force.

A similarity between the two theories is that they are both strongly affected by their modified expansion histories, and largely track the boost profile of their respective modified background-only boost curves. This reinforces one of the messages of \cite{Wright:2022krq}: that the background can play a significant, perhaps even dominant, role in large-scale structure formation, even if the deviations from \lcdm \ are relatively tame\footnote{Though it is worth remembering that one of the lessons from the work in this paper is that such statements can be frame-dependent! Arguably the Einstein-frame curve in the left panel of Fig.~\ref{fig:kmou_hubble_omega} is `tame', whilst the Jordan frame curve might be considered otherwise.}. 

\subsubsection{Redshift dependence}
\begin{figure}
    \centering
    \includegraphics[width=0.9\textwidth]{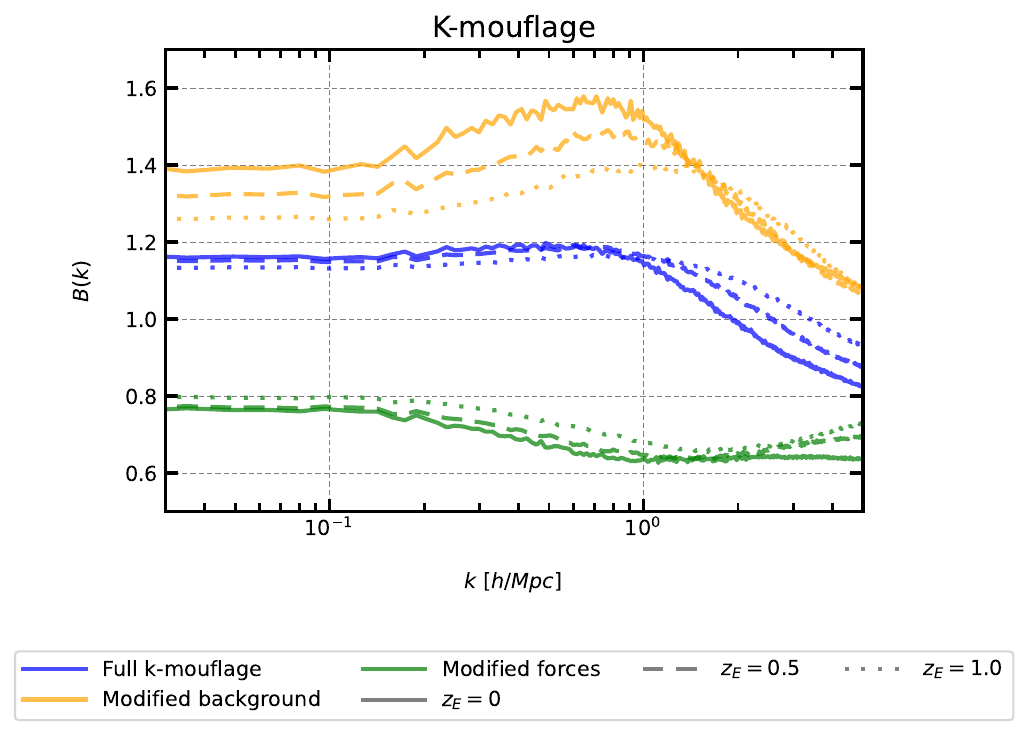}
    \caption{An exploration of the redshift dependence of the K-mouflage modified gravity effects presented in Fig.~\ref{fig:bk_breakdown}. $n=2$, $K_0=1, \ \beta_K=0.2$. Deviations from GR-\lcdm \ increase with passing time, tracking the behaviour of the scalar field as it becomes more significant at late times. The effect from the modified background shows the strongest evolution with redshift. The linear theory counterpart of this figure is the left panel of Fig.~\ref{fig:growth_breakdown}.}
    \label{fig:bk_kmou_breakdown_evolving_redshift}
\end{figure}

In Fig.~\ref{fig:bk_kmou_breakdown_evolving_redshift} we can see the breakdown of the K-mouflage boost at different redshifts. In general, we see that the deviation from GR-\lcdm \ grows with time, which tracks the behaviour of the scalar field. As the scalar field, and particularly its derivative, become significant at late times, the large-scale enhancement and small-scale suppression both increase. We can see that the effect from the modified background carries the strongest evolution with redshift\footnote{We can also see that despite the redshift evolution of the modified force effects looking comparatively weak, the net effect on the blue curve is small. One should not view the blue curve as being an average of the yellow and green curves, as the effects are not simply additive or multiplicative. For example, the conformal term does not only affect the gravitational forces of K-mouflage, but also affects the expansion history of the theory, owing to its presence in the background equations. The amplitudes of each component effect are best compared quantitatively amongst themselves, rather than with other effects.}. This is because the selected redshifts lie in the period when the K-mouflage Hubble rate is most different from that of \lcdm. That is, they pick out points on the trough in the left panel of Fig.~\ref{fig:kmou_hubble_omega}. The span of values of the ratio $H_{\textrm{kmou}} / H_{\Lambda\textrm{CDM}}$ is of relatively high magnitude, ranging from $22\%$ to $96\%$ of the magnitude of the local minimum at $z_E=0.13$ over these redshifts. This is likely when the transfer of power-like effects discussed in \cite{srinivasan_cosmological_2021} are most noticeable as a function of time. 

Meanwhile, we see that the curve with modified force effects has a comparatively weak evolution with redshift. This can partly be seen in Fig.~\ref{fig:linear_effects}. Over $z_E \in \{1.0,0.5,0.0\}$ we can see that the evolution over time of the coupling, $\mu$ is weak. Thus, on large scales, we see a weak evolution in the boost. The conformal factor maintains an appreciable evolution with time in this range, and so we see that on the quasi-linear and non-linear scales, when $\mu$ has diminished significance, there is a slight increase in the variation of the modified force boost curves over redshift.

It should be noted that one may expect to find non-linear contributions to the force that are not captured by a Vainshtein-style calculation of the perturbations to K-mouflage. However, based on the calculations of Section \ref{sec:Kmou_vs_Vainshtein}, such contributions can be expected to be on scales smaller than where the \textrm{COLA} approach has been observed to be accurate and where baryonic effects should be considered. The assumption of spherically symmetric configurations in the calculation used to derive the force is one of the key factors behind the limitations of \HiCOLA's accuracy on very small scales. See \cite{Brando:2023fzu} for an alternative approach that does not utilise the spherical approximation.

\subsection{Exploring the K-mouflage model space}
\subsubsection{Variation of $K_0,\beta_K$}\label{sec:evolving_k_b}

\begin{figure}
    \centering
    \includegraphics[width=1.0\textwidth]{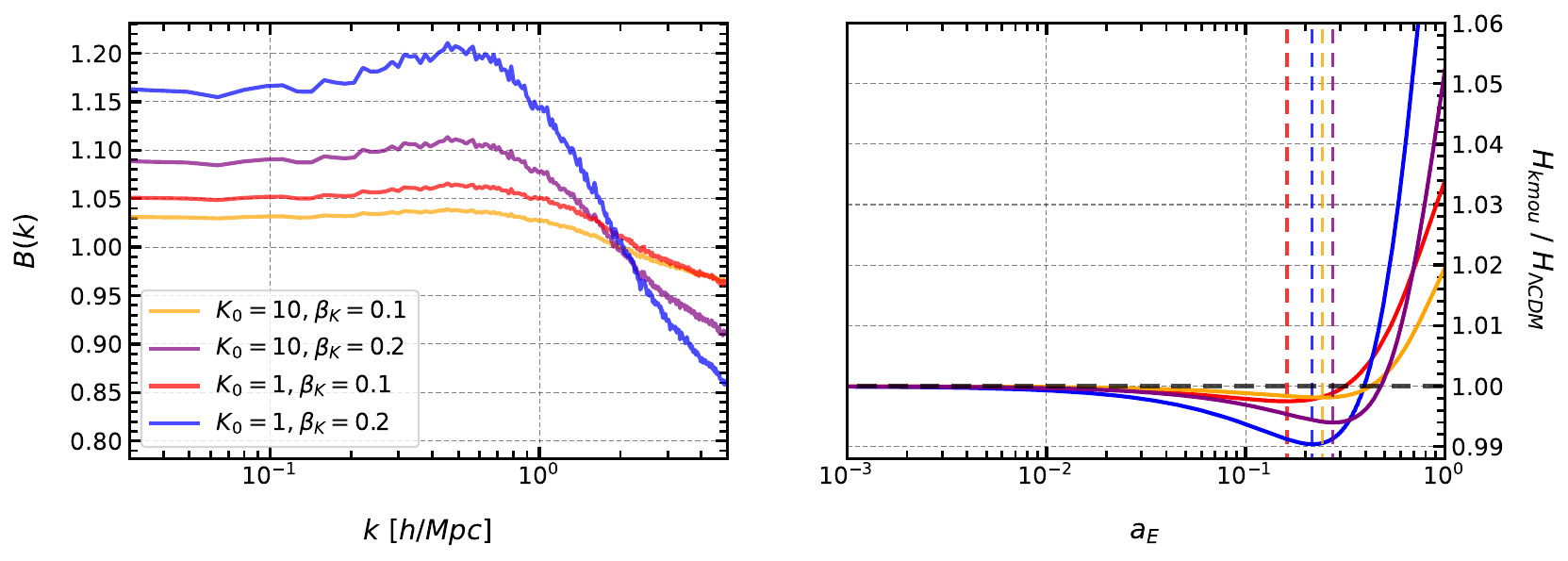}
    \caption{\textit{Left panel:} Comparison of the K-mouflage boost with respect to GR-\lcdm \ as $\{\beta_K, K_0\}$ [see eq.~\eqref{eq:conformal_factor},\eqref{eq:kmou_K}] are varied. The legend is shared with the right panel.\\
    \textit{Right panel:} K-mouflage Hubble rate in the Jordan frame relative to the Hubble rate of \lcdm \, for different values of the K-mouflage parameters (this is a multi-parameter counterpart of of the Jordan frame Hubble rate in the left panel of Fig.~\ref{fig:kmou_hubble_omega}). The legend is shared with the left panel. We see that a reduced value of $K_0$ and increased value of $\beta_K$ produces the most pronounced deviation from the Hubble rate of \lcdm \ (see the $K_0=1, \beta_K=0.2$ case in blue). As a result, the boost for the same model in the left panel of Fig.~\ref{fig:kmou_evolving_params} has the largest amplitude, as the enhancement coming from the background is strongest for this choice of the parameters. Increasing $\beta_K$ or $K_0$ also appears to shift the location of the minimum (dashed vertical lines) to later times, however it is unclear whether this is correlated with the location in scale of the maximum in the boosts. A larger set of parameters may be required to verify such a connection.}
    \label{fig:kmou_evolving_params}
\end{figure}

In the left panel of Fig.~\ref{fig:kmou_evolving_params} we can observe the effects of changing the K-mouflage parameters $\{K_0,\beta_K\}$, present in eqs.~\eqref{eq:kmou_K} and~\eqref{eq:conformal_factor}, on the boost. We see that raising $\beta_K$ in turn raises the large-scale amplification of structure formation in K-mouflage with respect to GR. For $K_0=1$, a doubling of $\beta_K$ results in an approximately $10.9\%$ increase in the boost. This is expected given the presence of $\beta_K$ in the coupling [eq.~\eqref{eq:kmou_mu}]. However, the precise effect of $\beta_K$ goes beyond a simple $1+2\beta_K^2$ contribution. $\beta_K$ also influences the expansion history of K-mouflage, and features in the conformal factor, which acts to suppress growth by the weakening of the gravitational force. For the former effect on expansion histories, we see that increasing $\beta_K$ leads to a minimum of greater magnitude in the relative Hubble rate, shown in the right panel of Fig.~\ref{fig:kmou_evolving_params}. This is therefore an additional effect that supports the increase of the amplitude of the boost with increased $\beta_K$. Meanwhile the conformal effect can be seen on the small scales, where increasing $\beta_K$ has led to increased suppression of structure relative to GR. Therefore, we find that the effect of $\beta_K$ in fact exacerbates the differences in clustering between large and small scales relative to GR.

We can see that the effect of increasing $K_0$ depends on the value that $\beta_K$ was held at. Comparing the blue and purple curves in the left panel of Fig.~\ref{fig:kmou_evolving_params} shows that decreasing $K_0$ has a similar effect on the boost to increasing $\beta_K$ but to a lesser extent, when $\beta_K=0.2$. The anti-correlation between $K_0$ and the enhancement of the boost on large scales is once again expected, as the coupling has a reciprocal relationship with $K_0$. The coupling, defined in eq.~\eqref{eq:kmou_mu}, becomes suppressed with increasing $K_0$. Meanwhile, comparing the red and yellow curves in the left panel Fig.~\ref{fig:kmou_evolving_params} when $\beta_K$ is at the lower value of $0.1$ indicates that the effect of $K_0$ is mostly visible on large scales, with little effect on small scales.  The pair-wise trends for $K_0$ in the left panel of Fig.~\ref{fig:kmou_evolving_params} suggest that $K_0$'s effects on small scale boost signal are in some sense controlled by $\beta_K$. When $\beta_K$ is small, the effects of $K_0$ are limited to the large scales. With increased values of $\beta_K$, the effect of $K_0$ on the small scales also rises in tandem.

\subsubsection{Variation of $n$}\label{sec:evolving_n}

\begin{figure}
    \centering
    \includegraphics[width=0.8\textwidth]{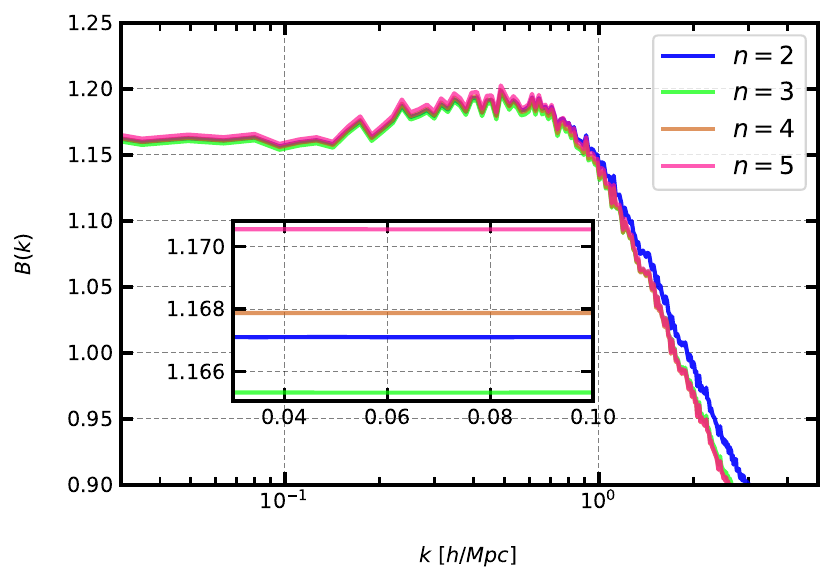}
    \caption{Effect of varying K-mouflage parameter $n$ [see eq.~\eqref{eq:kmou_K}] on the boost at $z_E=0$. The inset shows the scale-independent large-scale behaviour of the curves. The inset curves are computed using the linear power spectrum, and thus they lack the sample variance noise of their non-linear counterparts.}
    \label{fig:kmou_boost_evolvingn}
\end{figure}

In Fig.~\ref{fig:kmou_boost_evolvingn} we examine the effects of changing $n$ on the phenomenology of K-mouflage theories. As a reminder, as can be seen in eq.~\eqref{eq:kmou_K}, varying $n$ changes the order of $K$. As discussed in Section~\ref{sec:constraints}, we restrict our analysis to $n \in \mathbb{N}$, and fix $K_0 = 1$ and $\beta_K = 0.2$, as we saw that for $n=2$, the boost signal was strongest with these values.

Varying $n$ has a modest effect on the boost, appearing to mirror the effect of varying $\beta_K$, but to a lesser degree. On large scales, increasing $n$ appears to have sub-percent effects (see inset of Fig.~\ref{fig:kmou_boost_evolvingn}), and it is therefore difficult to ascertain any trends with certainty. On small scales, an increase in $n$ results in further suppression, relative to GR-\lcdm. This is driven by the behaviour of the conformal factor, $A$. As $n$ is increased, the conformal factor reaches a lower global minimum at $a_E=1$, thereby leading to weaker gravitational forces between particles. The effect of $n$ on the boost is far less significant than those of $K_0$ and $\beta_K$ for the values studied in this paper.

\section{Convergence tests}
\label{sec:convergence}
As justified in Section~\ref{sec:Kmou_vs_Vainshtein}, our approach to simulate K-mouflage gravity in the Jordan frame neglects the effect of screening. Hence its only limitations (besides possible inaccuracies due to lack of screening) are due to the finite resolution employed in our simulations. In this section we present a study of the impact of time, force, and mass resolution on the boost factor with a focus on the deep non-linear regime. 

We focus on the $n=2$, $K_0=1$, $\beta_K=0.2$  model and run a high-resolution simulation in K-mouflage and its $\Lambda$CDM counterpart using the same box size as for the other simulations discussed in Section~\ref{sec:power_spectra} ($L=400\mpcoh$), but with an increased number of particles and force mesh ($\NP=1024$, $\NM=3072$) and with a larger number of time steps ($\NS=100$). These new simulations thus have $8$ times the mass resolution ($\MP \approx 5.1 \cdot 10^{9} \Msun$), approximately $2$ times the force resolution $\LF \approx 0.13 \mpcoh$, and $2.5$ times the time resolution $\Delta a \approx 0.01$ of the simulations discussed in previous sections. We use the boost factors computed from these high-resolution simulations as \textit{references} to study the sensitivity of the K-mouflage boost factor to the resolution parameters by comparing with simulations with default mass resolution ($\MP \approx 4.1 \cdot 10^{10} \Msun$). In these default mass resolution simulations, we vary the force and time resolution parameters separately.

\begin{figure}
    \centering
    \includegraphics[width=\textwidth]{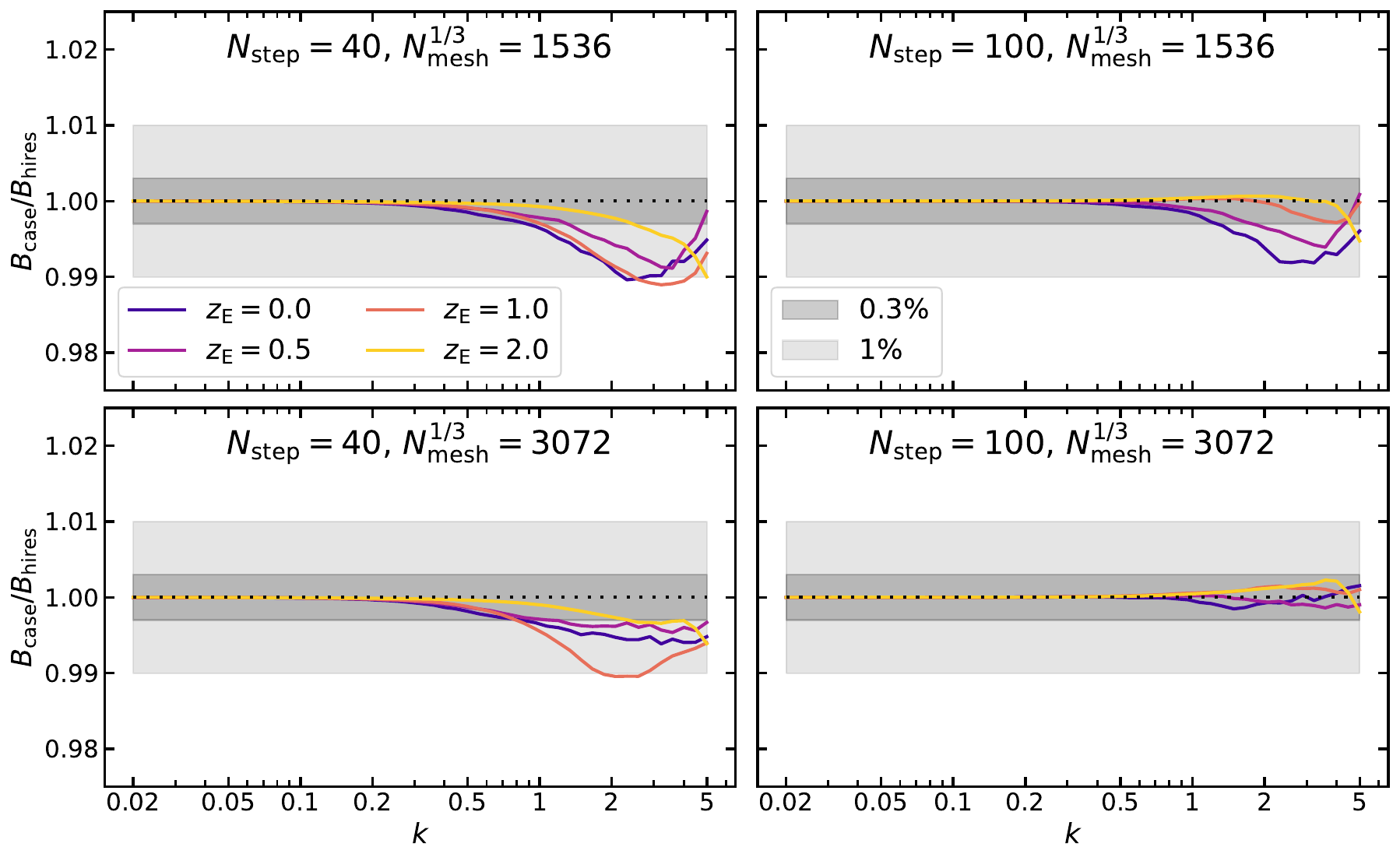}
    \caption{Ratio of matter power spectrum boost factors for the K-mouflage model $n=2$, $K_0=1$, $\beta_K=0.2$, computed with increasing force (top to bottom) and time (left to right) resolution settings with respect to the reference ones computed from the high-resolution simulations.}
    \label{fig:convergence}
\end{figure}

In Fig.~\ref{fig:convergence} we show the impact of force and time resolution on the predicted K-mouflage boost factors. 
Firstly, from the top left panel we notice that our default settings provide boost factors that are in approximately $1\%$ (approximately $0.3\%$) agreement up to $k=5 \hompc$ ($k=1 \hompc$), which is already a reassuring result. However, to get higher accuracy, it is not enough to increase only the force resolution or only the time resolution, as can be seen by observing that the bottom-left and top-right panels are characterised by the same level of convergence as the top-left panel. Instead, increasing both force and time resolution simultaneously provides a better level of convergence towards the high-resolution result, with deviations of roughly less than $0.3\%$ up to $k=5 \hompc$, as shown in the bottom-right panel. Furthermore, since in the bottom-right panel the force and time resolution settings are the same as the reference simulations, the only difference being the mass resolution, we can estimate that increasing the mass resolution above $\MP \approx 4.1 \cdot 10^{10} \Msun$ has a negligible effect on the K-mouflage boost factors in this range of scales and redshifts. 

\begin{figure}
    \centering
    \includegraphics[width=0.8\textwidth]{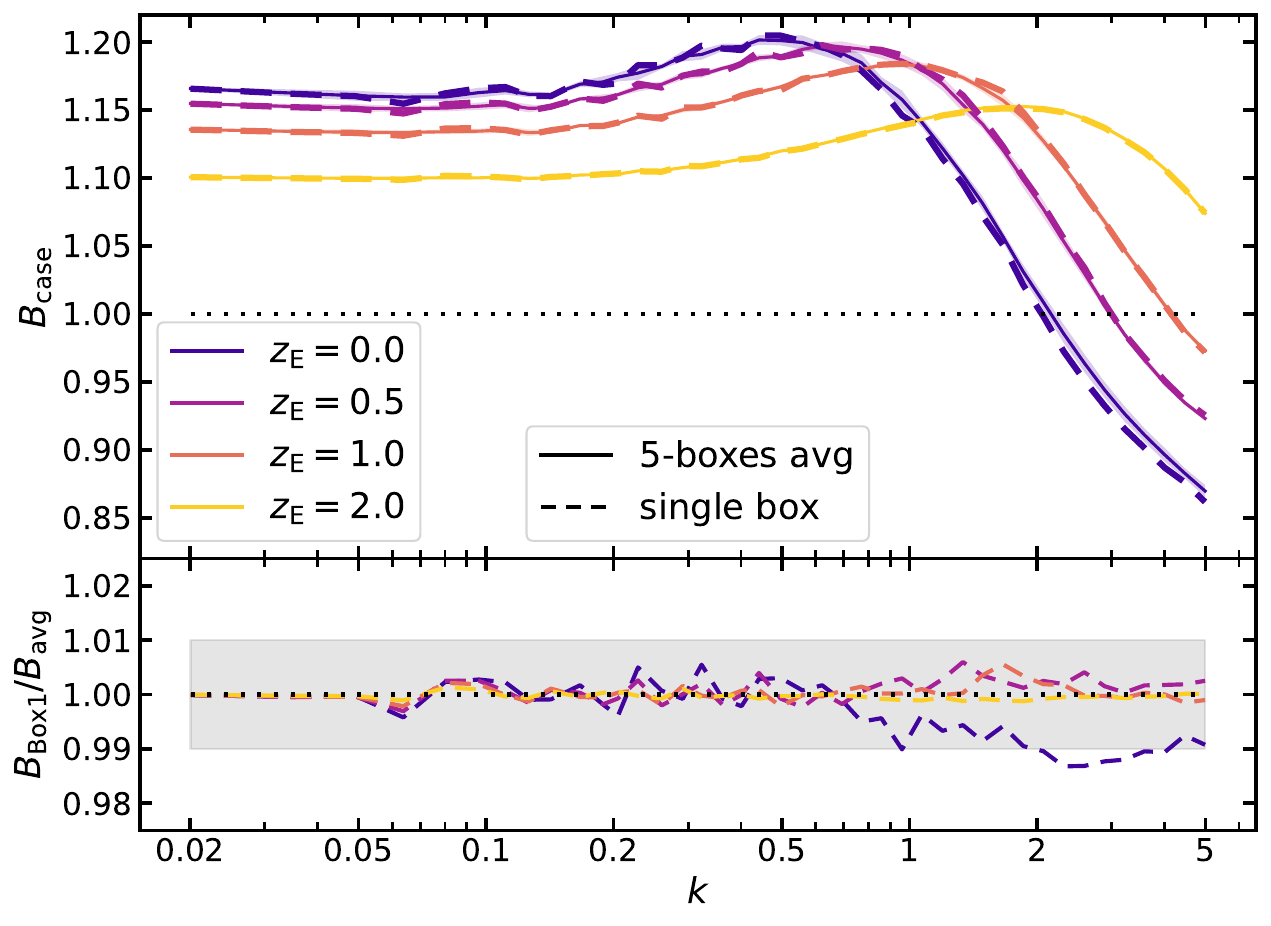}
    \caption{Comparison of the matter power spectrum boost factors for the $K$-mouflage model $n=2$, $K_0=1$, $\beta_K=0.2$, computed with a single realisation with respect to one computed averaging over 5 realisations.}
    \label{fig:sample_variance}
\end{figure}

A possible additional source of error in our predictions of the boost factor is the finite volume used for our simulations. To investigate the importance of this effect we run 5 realisations of the default resolution simulations and compare the single realisation used for all the results presented so far against the average over 5 realisations. The results of this exercise are summarised in Fig.~\ref{fig:sample_variance}, where we can see that sample variance is responsible for $\mathcal{O}(1)\%$ inaccuracies of the single-box predictions only at $z=0$ and for $k \gtrsim 1 \hompc$, while the error is negligible for higher redshift values or smaller wave-numbers.

\section{Conclusions}
\label{sec:conclusions}

The inclusion of K-mouflage in \HiCOLA is a key expansion in the conquest to constrain gravity theories with screening mechanisms using LSS. Whilst the K-mouflage models considered in this work do not exhibit screening behaviour on scales relevant to our simulations, we have seen that the combination of their  linear-regime behaviour and the effects of the conformal factor give rise to non-trivial small-scale dynamics visible in the boost. The small-scale departures in clustering from that of GR-\lcdm{} are driven by the competition between the modified background, and the weakening of the gravitational force by the conformal term. We have also seen that $\beta_K$, the coefficient of the exponent of $A$ (eq.~\eqref{eq:conformal_factor}), most strongly affects the shape of the power spectrum. In contrast, and perhaps surprisingly, changing the order of the polynomial form for $K$ has modest effects.

The implementation of K-mouflage in \HiCOLA{} also carries the distinction of being a rapid predictor of K-mouflage clustering results in the Jordan frame. The conformal term of K-mouflage brings with it the added needed to pay heed to the frame in which results were computed to enable proper comparisons. We have seen that background quantities, such as the Hubble rate and fractional energy densities, can look significantly different between the Jordan and Einstein frames as these are not frame-independent observables. Conversely, we also saw that the linear growth and power spectrum, though also not observables, do not transform between the frames. This is the result of a cancellation of the contributions from the transformed Hubble rates, and the contributions of the conformal term. However, even in comparing power spectra, care must be taken. The transformation of redshifts between frames must be taken into account to ensure the comparison of the correct power spectrum from each frame.

We have also seen that the lack of hierarchy in the contributions of perturbations preclude a derivation for a generic fifth force with K-mouflage screening effects, as done for Vainshtein screening in \cite{Wright:2022krq}. However, we also saw that on scales of $k \sim \mathcal{O}(1)$ h/Mpc, that a Vainshtein-like treatment is sufficient for K-mouflage theories, as K-mouflage screening is heavily suppressed on the scales accessible to COLA simulations.  

Armed with this knowledge, we have therefore successfully implemented K-mouflage theories with polynomial kinetic terms in \HiCOLA{}, providing clustering predictions in the Jordan frame. The comparison of \HiCOLA's results with other codes, including Einstein-frame implementations and $N$-body codes, is the subject of \cite{Bose:2024qbw}. The scope of \HiCOLA{} as a tool to address the Horndeski class at large has thus widened, though we note that Chameleon screening remains to be incorporated and we leave this for future work. Being able to produce rapid clustering predictions for K-mouflage theories using \HiCOLA{} opens the door to generating training data for K-mouflage emulators. These emulators need not be limited to predicting summary statistics, as recent interest has been sparked in field-level predictors~\cite{Saadeh:2024vuj}. This will enable constraints of the K-mouflage theory space using large-scale structure observations, serving as a complement to existing methods and possibly probing new regions of the K-mouflage parameter space.

\acknowledgments

We would like to thank Prof. Tsutomu Kobayashi for the valued discussions and guidance on comparing K-mouflage and Vainshtein screening scales. We are grateful to Dr Ben Bose for providing us with early background solution data which we could use as reference. We also thank Dr Guilherme Brando for the discussions that helped in our understanding of conformal transformations. A.S.G. is supported by a STFC studentship. B.F. is supported by Royal Society grant no. RF$\backslash$ERE$\backslash$210304. T.B. is supported by ERC Starting Grant SHADE (grant no.\,StG 949572) and by a Royal Society University Research Fellowship (grant no.\,URF$\backslash$R$\backslash$231006). \HiCOLA{} data was generated utilising Queen Mary's \href{https://doi.org/10.5281/zenodo.438045}{Apocrita High-Performance Computing (HPC) facility}~\cite{QMUL_apocrita}, supported by QMUL Research IT. Numerical computations were also done on the Sciama HPC cluster, which is supported by the Institute of Cosmology and Gravitation, SEPNet and the University of Portsmouth.

\bibliographystyle{JHEP}
\bibliography{References}


\appendix

\section{Comparisons of significance for perturbations to the Horndeski action}
\label{sec:Kmou_vs_Vainshtein_appendix}

In this section we include the calculations that substantiate the points made in Section~\ref{sec:Kmou_vs_Vainshtein}.

Let us give an example of what we had meant by ``comparable to each other''. Wanting, say, the first two terms in eq.~\eqref{eq:galileon_action} to be comparable to each other entails

\begin{equation}
    M^2_P R \sim X|_{\textrm{FLRW}} = \frac{\dot{\bar{\phi}}^2}{2}.
\end{equation}
$\bar{\phi}$ is notation for the scalar field, $\phi$, evaluated on an FLRW background. 

We consider perturbations about an FLRW background in the Newtonian gauge. That is, 
\begin{equation}
    ds^2 = -(1+2\Phi)dt^2 + (1+2\Psi)d\vec{\textbf{x}}^3
\end{equation}
and 
\begin{equation}
    \phi = \bar{\phi} + \pi,
\end{equation}
where $\pi$ is the \textit{dimensionless} perturbation about the FLRW background value, $\bar{\phi}$, and is small. We consider these perturbations to be sourced by non-relativistic matter, with a stress energy defined entirely by $T^{\mu}{}_{\mu} \sim -\rho$. As we are interested in scales much smaller than both the Hubble horizon and the horizon of the scalar field, we can apply the quasi-static approximation (QSA)\footnote{Strictly speaking, applying the QSA is valid provided the scalar sound speed is close to lightspeed, $c_{\phi} \sim 1$.}~\cite{Noller:2013wca,Sawicki:2015zya,Pace:2020qpj}. This means we assume time derivatives on perturbations are negligible and only consider spatial derivatives. In short, assuming an FLRW background implies and the QSA implies $\partial_i \bar{\phi} = 0$ ($i$ is used for spatial indices), and $\partial_t \pi = 0$,  $\partial_i \pi \gg \partial_t \bar{\phi}$ respectively on the sub-horizon scales we consider.

We can expand the terms in the action~\eqref{eq:galileon_action} in $\pi$ to obtain the perturbed action. We focus on $\mathcal{O}(\pi^2)$ and higher, as the linear-$\pi$ term gives the background equations, presented in Section~\ref{sec:kmou_bg_einstein}. $\partial$ denotes a spatial derivative. 

\underline{From the $X^m/\Lambda^{4m-4}$} (`K-mouflage') set of terms we find:

\begin{equation}
    M_P^2 \left[ (\partial \pi)^2 + \frac{(\partial \pi)^4}{H_0^{2}} + ... + \frac{(\partial \pi)^{2m}}{H_0^{2m-2}} + ... \right].
\end{equation}

\underline{From the $X (\Box \phi)^p / \Lambda_3^{3p}$} (`Vainshtein') set of terms we find:

\begin{equation}
   M_P^2 \left[ \frac{(\partial \pi)^2 (\partial^2 \pi)^2}{H_0^4} + ... + \frac{(\partial \pi)^2 (\partial^2 \pi)^{2p}}{H_0^{2p}} +...\right].
\end{equation}
The other terms in eq.~\eqref{eq:galileon_action} can be expanded in $\pi$ about $\bar{\phi}$ similarly.

At leading order one considers the contribution from $(\partial \pi)^2$. All other terms are considered insignificant and dropped. As a result, when combined with the matter action, one obtains the scalar field equation of the form (we write $\sim$ as we drop any overall numerical prefactors):

\begin{equation}
    M_P^2 \partial^2 \pi \sim \rho,
\end{equation}
which we can identify as Poisson's equation, and has the solution
\begin{equation}\label{eq:perturb_linear_soln}
    \pi \sim \frac{r_g}{r}
\end{equation}
where
\begin{equation}
    r_g = G_N M.
\end{equation}
$M$ is the mass of the source giving rise to the spacetime curvature and scalar field profile. As we will utilise the spherical approximation, the source is the matter distribution arranged in a sphere of constant density (a ``top hat" density profile). Hence, we we can write in terms of the coordinate $r$ exclusively, and $\partial \rightarrow \partial_r$.

We can now ask ourselves when a situation may arise where the other, higher order, terms in this perturbative expansion are comparable to the quadratic term. For example, when is $H_0^{-2} (\partial \pi)^4$, coming from the $X^2$ term, as significant as $(\partial \pi)^2$, the linearised term? This occurs when

\begin{align}
    (\partial \pi)^2 &\lesssim H_0^{-2} (\partial \pi)^4 \\
   \implies 1 &\lesssim  H_0^{-2} (\partial \pi)^2
\end{align}
We can utilise the solution for $\pi$ that we get from the quadratic term in eq.~\eqref{eq:perturb_linear_soln} as an ansatz here, giving
\begin{align}
    1 &\lesssim H_0^{-2} \left( \partial_r \left[ \frac{r_g}{r}\right] \right)^2\\
    \implies 1 &\lesssim H_0^{-2} \left( -\frac{r_g}{r^2} \right)^2\\
    \implies r^4 &\lesssim H_0^{-2} r_g^2\\
    \implies r &\lesssim\sqrt{\frac{r_g}{H_0} }.
\end{align}
Let us define
\begin{equation}
\label{eq:rX2_app}
     r_{X^2} \coloneqq \sqrt{\frac{r_g}{H_0} }.
\end{equation}
What we have found is that when we are at radial distances of below $r_{X^2}$ from the matter source, the perturbation coming from $X^2$ becomes as significant as the quadratic term and should not be neglected.

What about when $H_0^{-2} (\partial \pi)^2 \partial^2 \pi$ , coming from $X \Box \phi / \Lambda_3^3$, is comparable to the linearised term? That is,
\begin{align}
    (\partial \pi)^2 &\lesssim H_0^{-2} (\partial \pi)^2 \partial^2 \pi\\
    \implies 1 &\lesssim H_0^{-2}\partial^2 \pi\\
    \implies 1 &\lesssim H_0^{-2} \frac{2r_g}{r^3}\\
    \implies r &\lesssim r_V \coloneqq \left( \frac{2 r_g}{H_0^2} \right)^{1/3}.
\end{align}
Notice that $r_V > r_{X^2}$ is true if
\begin{equation}\label{ineq:rv_vs_rx}
    r_g < \frac{4}{H_0}.
\end{equation}

Remember that we work in units where $c=1$ and that $r_g$ is of the order of the Schwarzschild radius for an object. Therefore, the inequality amounts to a comparison between the Hubble radius and the Schwarzschild radius for a given object, and indeed for galactic clusters the former is much larger than the latter.

This implies there is a region, $r_{X^2} < r < r_V$, where  $H_0^{-2} (\partial \pi)^2 \partial^2 \pi$ is comparable to the quadratic term, but  $H_0^{-2} (\partial \pi)^4$ is not and can be neglected. What one finds is that a similar situation occurs for any of the terms arising from  $X^m/\Lambda^{4m-4}$; there is always a regime, defined by the Vainshtein radius $r_V$ where these terms are subdominant and can be neglected, while  $H_0^{-2} (\partial \pi)^2 \partial^2 \pi$ is kept. 
We can see this by examining the general term that goes as $X^m (\Box \phi)^p /(\Lambda_2^{4m-4} \Lambda_3^{3p})$. This is comparable to the linearised term from $X$ when
\begin{equation}
    r^{4m + 3p - 4} \lesssim \frac{ 2^p r_g^{2m+p-2}}{H_0^{2(m+p-1)}} .
\end{equation}
We can see that if $p=0$, restricting ourselves to the K-mouflage terms, then we find that the $m$ dependence drops out:
\begin{align}
    r^{4m-4} &\lesssim \frac{r_g^{2m-2}}{H_0^{2m-2}}\\
    \implies r &\lesssim \sqrt{\frac{r_g}{H_0}}\\
    \implies r &\lesssim r_{X^m} \coloneqq \sqrt{\frac{r_g}{H_0}}.
\end{align}
Note this is the same scale as derived in eq.\eqref{eq:rX2_app}. This means that all $X^m$ terms in eq.~\eqref{eq:galileon_action} are as significant as the quadratic term on the same scales.

This is why it is sufficient to treat K-mouflage theories by considering only their behaviour in this intermediate regime where Vainshtein behaviour dominates for \HiCOLA{} results. We can use the expression for the screening factor derived in \cite{Wright:2022krq} for K-mouflage. The screening factor is appearing in the \HiCOLA{} fifth force expression~\eqref{eq:fifth_force_gg4} is:

\begin{align}
    S&=\frac{2}{\chi}\left(\sqrt{1+\chi} -1\right). \label{eq:screen_coeff}
\end{align}
where
\begin{align}
\chi &= \frac{{\cal B}{\cal C} \,\Omega_{\rm m0}}{E a^3}~\frac{G_{G_4}}{G_{\rm N}} \delta_{\rm m}\label{eq:chioverdelta}
\end{align}
and $\cal B,\,C$ are defined in section 3 of \cite{Wright:2022krq}. Using the K-mouflage Horndeski functions in eq.~\eqref{eq:kmou_J}, and the choices for $K$ and $A$ defined in eqs.~\eqref{eq:kmou_K}, \eqref{eq:conformal_factor} respectively, results in ${\cal B}=0$ and hence $S \rightarrow 1$. This is because of the lack of $X$ dependence in $G_3$, i.e. because $G_{3X} = 0$. This is consistent with the result of \cite{brax_K-mouflage_2014_LSS}: that K-mouflage theories do not exhibit screening on mildly non-linear scales.

\end{document}